\documentclass[a4paper,11pt]{article}
\pdfoutput=1 

\usepackage{jheppub} 

\usepackage{bm}

\title{\boldmath Superconformal blocks from Wilson lines with loop corrections}

\author[a]{Yasuaki Hikida}
\author[b]{Takahiro Uetoko}

\affiliation[a]{Center for Gravitational Physics, Yukawa Institute for Theoretical Physics, Kyoto University,\\ Kyoto 606-8502, Japan}

\affiliation[b]{Department of Physical Sciences, College of Science and Engineering, Ritsumeikan University,\\Shiga 525-8577, Japan}

\emailAdd{yhikida@yukawa.kyoto-u.ac.jp}
\emailAdd{rp0019fr@ed.ritsumei.ac.jp}

\abstract{We compute the $\mathcal{N}=1$ superconformal blocks  from  the networks of open Wilson lines in the $\text{osp}(1|2)$ Chern-Simons theory  in the expansion of large central charge $c$. We first reproduce the $1/c$ correction of conformal weight from an open Wilson line by adopting the regularization prescription developed in our previous works.
	We then obtain the closed form expressions of superconformal blocks including $1/c$ corrections, which were not available before. We also examine heavy operators corresponding to supersymmetric conical spaces, and the geometry is quantized by utilizing the coadjoint orbits of the super Virasoro group. Superconformal blocks involving these operators are also examined.}

\arxivnumber{1806.05836}
\preprint{YITP-18-64}

\begin{document}
	\maketitle
	\flushbottom

\section{Introduction}

In \cite{Hikida:2017ehf,Hikida:2018dxe}, we computed the conformal blocks of the 2d W$_N$ minimal model from the bulk viewpoints in the large $c$ expansion, where $c$ is the central charge of the model. The large $c$ regime of the model is supposed to be described by the 3d sl$(N)$ Chern-Simons gauge theory \cite{Gaberdiel:2010pz,Castro:2011iw,Gaberdiel:2012ku,Perlmutter:2012ds}, and
it was proposed that 
the conformal blocks can be computed from the networks of open Wilson lines in the Chern-Simons theory \cite{Verlinde:1989ua,Bhatta:2016hpz,Besken:2016ooo}.
It is notoriously difficult to deal with quantum gravity effects generically.
Nevertheless, we were able to offer a prescription to regularize divergences from loop diagrams associated with the Wilson line networks by making use of the boundary symmetry. Moreover, conformal blocks are known to be useful ingredients, e.g., for conformal bootstrap programs (see \cite{ElShowk:2012ht} for instance).
Our previous works confirmed that the Wilson line method is a promising approach to compute conformal blocks in $1/c$ expansion. Related works may be found in \cite{Fitzpatrick:2016mtp,Besken:2017fsj,Anand:2017dav,Bhatta:2018gjb}.

In this paper, we consider a supersymmetric extension of the previous works.
There are two main motivations for introducing supersymmetry.
The first one is on the suppression of quantum effects and the second one is on the relation to superstring theory. We mainly focus on the $\mathcal{N}=1$ superconformal blocks in this paper, but 
we can discuss  extensions to a theory with $\mathcal{N}=2$ W$_{N+1}$ symmetry. Its large $c$ regime is supposed to be described by the sl$(N+1|N)$ Chern-Simons gauge theory \cite{Creutzig:2011fe,Hikida:2012eu}, and 
there seem suppressions in the $1/c$ corrections of conformal weights, for instance.
The relation between higher spin theory and superstring theory can be also argued with extended supersymmetry. As concrete examples, there are proposals on higher spin holographies with $\mathcal{N}=3$ supersymmetry in \cite{Creutzig:2013tja,Creutzig:2014ula,Hikida:2015nfa} and with $\mathcal{N}=4$ supersymmetry in \cite{Gaberdiel:2013vva,Gaberdiel:2014cha}. One of the aim of this paper is to prepare for the analysis on these complicated cases.

As a simple and specific example, we examine the 2d $\mathcal{N}=1$ super Virasoro minimal model \cite{Friedan:1984rv} from the bulk viewpoints.
We examine the superconformal blocks of  the model from the 3d osp$(1|2)$ Chern-Simons gauge theory \cite{Achucarro:1987vz} in the large $c$ expansion.
We first examine the expectation value of an open Wilson line with divergences from loop diagrams regularized by adopting the prescription in \cite{Hikida:2017ehf,Hikida:2018dxe}.
From this computation, we read off the $1/c$ corrections to the conformal weight $h$ of light operator with $h = \mathcal{O}(c^0)$.
We then compute superconformal blocks involving these light operators from the networks of open Wilson lines including the $1/c$ corrections. We evaluate the identity four point blocks up to the $1/c^2$ order, and consider three and general four point blocks up to the $1/c$ order.%
\footnote{The four point blocks with the identity (general) operator exchanged are called as the identity (general) blocks.}
In particular, we find the closed form expressions of the four point blocks with $1/c$ corrections. We then analyze the $1/c$ corrections for heavy operators with $h = \mathcal{O}(c)$, which correspond to supersymmetric conical spaces. For the purpose, we extend the analysis for the bosonic case in \cite{Raeymaekers:2014kea}, which uses the quantization of the coadjoint orbits of the Virasoro group, see \cite{Witten:1987ty,Alekseev:1988ce}. 
We further evaluate a heavy-heavy-light-light block with $1/c$ corrections from an open Wilson line in a supersymmetric conical space. See \cite{Fitzpatrick:2014oza,Chen:2016cms,Cornagliotto:2017dup,Alkalaev:2018qaz} for some recent works on the large $c$ limit of $\mathcal{N}=1$ superconformal blocks.

The organization of this paper is as follows.
In the next section, we summarize several basics of the $\mathcal{N}=1$ superconformal field theory. We first introduce the minimal model of  $\mathcal{N}=1$ super Virasoro algebra and then study superconformal blocks. In section \ref{Wilson}, we introduce an open Wilson line in the osp$(1|2)$ Chern-Simons gauge theory and explain our prescription to regularize divergences arising from loop diagrams. Applying the prescription, we reproduce the $1/c$ corrections in the conformal weight of light operator. In section \ref{superCB}, we compute the $1/c$ corrections in the three and four point blocks from the networks of open Wilson lines. In particular, we obtain the closed form expressions for the four point blocks including $1/c$ corrections.
In section \ref{heavy}, we examine the $1/c$ correction of heavy operator from the supersymmetric conical defect geometry in the osp$(1|2)$ Chern-Simons gauge theory. We first compute the $1/c$ correction of the conformal weight by quantizing the coadjoint orbits of the super Virasoro group. We then study a heavy-heavy-light-light block from an open Wilson line in the supersymmetric conical space. Section \ref{conclusion} is devoted to conclusion and discussions.
In appendix \ref{N1WT}, we include the correlators of superconformal currents  and give a derivation of a superconformal Ward-Takahashi identity.
In appendix \ref{4ptblocks}, we obtain the large $c$ limit of four point blocks and the expressions of four point blocks with degenerate operators in the expansion of $z$ and $1/c$ with the techniques of superconformal field theory.

\section{Basics of $\mathcal{N}=1$ superconformal field theory}
\label{Basics}

We start with a review on the $\mathcal{N}=1$ super Virasoro minimal model.
The $\mathcal{N}=1$ superconformal algebra is generated by energy-momentum tensor $T(z)$ with dimension 2 and superconformal generator $G(z)$ with dimension $3/2$.
In terms of mode expansions, the (anti-)commutation relations are
\begin{align}
\begin{aligned}
&[L_m , L_n] = (m-n) L_{m+n} + \frac{c}{12} m(m^2 -1) \delta_{m+n} \, , \\
&[L_m , G_r] =  \left( \frac{m}{2} - r \right) G_{m+r} \, , \quad
\{ G_r , G_u \} = 2 L_{r + u} + \frac{c}{3} \left( r^2 - \frac14 \right) \delta_{r+u}\, ,
\end{aligned}
\label{commutators}
\end{align}
where $r , u \in \mathbb{Z} + 1/2$ for the NS-sector and $r , u \in \mathbb{Z}$ for the R-sector.
A superconformal primary state $|h \rangle$ satisfies
\begin{align}
L_0 | h \rangle = h | h \rangle \, , \quad
L_n | h \rangle = G_r | h \rangle = 0 
\label{Primary}
\end{align}
with $n,r > 0 $.

The unitary minimal models are defined with the discrete values of central charge 
\begin{align}
c = \frac{3}{2} - \frac{12}{(k+2)(k+4)}
\label{center}
\end{align}
with $k=0,1,2,\cdots$.
The conformal weight of primary state is \cite{Friedan:1984rv}
\begin{align}
h_{m ,n} = \frac{((k+4)m - (k+2) n)^2 - 4 }{8 (k+2)(k+4)} + \frac{1 - (-1)^{m+n}}{32} 
\label{hmn}
\end{align}
with $m+n \in 2 \mathbb{Z}$ for the NS-sector and  $m+n \in 2 \mathbb{Z}+1$ for the R-sector. Here we are interested in an analytic continuation of the minimal models with large $c$. The expression of central charge in \eqref{center} suggests that%
\footnote{We can take $k = - 2 + \mathcal{O}(c^{-1})$, but in that case we should replace $m$ and $n$ in \eqref{hmn}.}
\begin{align}
k = -4  +\frac{6}{c} + \frac{27}{c^2} + \mathcal{O}(c^{-3}) \, .
\end{align}
The conformal weight in \eqref{hmn} is expanded in $1/c$ as
\begin{align}
\begin{aligned}
h_{m ,n} = & \left(\frac{1}{24}-\frac{n^2}{24}\right) c  +\frac{1}{16} \left(-4 m n+5 n^2-1\right) + \frac{1 - (-1)^{m+n}}{32}  \\ & -\frac{3 \left(m^2-n^2\right)}{8 c} -\frac{45 \left(m^2-n^2\right)}{16 c^2} + \mathcal{O}(c^{-3}) \, .
\end{aligned}
\end{align}
We first study light states with conformal weight of order $c^0$.
This means that we should set $n=1$. Rewriting $m = 4 j +1$, we have
\begin{align}
h_j = -j - \frac{3 j}{c}(2 j+ 1 ) - \frac{45 j}{c^2} \left( j + \frac{1}{2}  \right)+ \mathcal{O}(c^{-3}) \, .
\label{hexp}
\end{align}
We only consider the NS-sector for light states, thus $j \in \mathbb{Z}/2$.
We describe these states in terms of bulk open Wilson lines.
The conformal weights of other states are of order $c$, and we consider
a particular set of operators with $(s=1,2,3,\ldots)$
\begin{align}
\label{h1s}
h_{1 ,s} =\left(\frac{1}{24}-\frac{s^2}{24}\right) c  +\frac{1}{16} \left(5 s^2-4 s-1\right) + \frac{1 + (-1)^{s}}{32} + \mathcal{O}(c^{-2}) \, .
\end{align}
Here $s$ is odd for the NS-sector and $s$ is even for the R-sector.
These state are conjectured to corresponds to supersymmetric conical spaces.

In order to analyze correlation functions, it is convenient to work with superspace $(z , \theta)$, see, e.g., \cite{Friedan:1986rx,AlvarezGaume:1991bj}.
The generators of superconformal algebra are written in terms of a superfield $T (z , \theta) = T_F (z) + \theta T_B(z) $ with $T(z) = T_B (z)$ and $G(z) = 2 T_F (z) $.
For a superconformal primary, we define a superfield  as
\begin{align}
\Phi_h (z , \bar z , \theta , \bar \theta) = V_h (z , \bar z) +\theta  \Lambda_h  (z , \bar z) + \bar \theta \bar{\Lambda}_h (z , \bar z) + \bar \theta  \theta W_h (z , \bar z)\, ,
\label{superfield}
\end{align}
where
\begin{align}
\Lambda_h = G_{-1/2} V_h \, , \quad 
\bar \Lambda_h = \bar G_{-1/2} V_h \, , \quad
W_h = G_{-1/2} \bar G_{-1/2} V_h \, .
\end{align}
In terms of superfield, the two point function can be written as
\begin{align}
\langle \Phi_{h} (z_1 , \theta_1) \Phi_{h} (z_2 , \theta_2)  \rangle = \frac{1}{|Z_{12}|^{4 h}} \, , \qquad
Z_{ij} = z_i - z_j - \theta_i \theta_j \, .
\label{2ptsuper}
\end{align}
With component fields, the two point functions are
\begin{align}
\langle V_{h} (z_1 ) V_{h} (z_2 )  \rangle = \frac{1}{|z_{12}|^{4 h}}  \, , \quad
\langle W_{h} (z_1 ) W_{h} (z_2 )  \rangle = - \frac{(2 h)^2}{|z_{12}|^{4 h + 2}} \, .
\label{2ptcomp}
\end{align}
Expanding the conformal weight in $1/c$ as
\begin{align}
h = h^{(0)} + \frac{1}{c} h^{(1)} + \frac{1}{c^2} h^{(2)} + \mathcal{O} (c^{-3}) \, , 
\end{align}
the two point function becomes
\begin{align}
\begin{aligned}
&\langle V_{h} (z_1 ) V_{h} (z_2 )  \rangle  
\\& =\frac{1}{z^{2h^{(0)}}}\left[ 1-\frac{2h^{(1)}\log(z)}{c} + \frac{(2 (h^{(1)} \log(z) )^2 -2h^{(2)}\log(z))   }{c^2} \right] + \mathcal{O}(c^{-3}) \, ,
\end{aligned}
\label{VVexp}
\end{align}
where only the holomorphic sector is expressed.
We can also expand $\langle W W \rangle$ in $1/c$ in a similar manner.

The three point function of superfields depends on $(3|3)$ supercoordinates,%
\footnote{Here $(n|m)$ means the $n$ bosonic and $m$ fermionic degrees of freedom.}
while only $(3|2)$ parameters can be fixed by using osp$(1|2)$ subalgebra as the global part of super Virasoro algebra. Therefore, the three point function is given by a function of a fermionic superconformal invariant
\begin{align}
\theta_{123} = (Z_{12} Z_{13} Z_{32} )^{-1/2} (Z_{23} \theta_1 + Z_{31} \theta_2 + Z_{12} \theta_3 + \theta_1 \theta_2 \theta_3) \, .
\end{align}
For spinless operators, we thus find
\begin{align}
\langle \Phi_{h_1} (z_1 , \theta_1) \Phi_{h_2} (z_2 , \theta_2) \Phi_{h_3} (z_3 , \theta_3) \rangle = \frac{C_{123} + | \theta_{123} |^2  \tilde C_{123} }{|Z_{12}|^{2 h_{12}}  |Z_{13}|^{2 h_{13}}|Z_{32}|^{2 h_{32}}}
\label{3ptsuper}
\end{align}
with $h_{12} = h_1 + h_2 - h_3$ and so on. In particular, we have
\begin{align}
&\langle V_{h_1} (z_1 ) V_{h_2} (z_2) V_{h_3} (z_3 ) \rangle \propto \frac{1 }{|z_{12}|^{2 h_{12}}  |z_{13}|^{2 h_{13}}|z_{32}|^{2 h_{32}}}\, ,
\label{VVV}
\end{align}
and
\begin{align}
&\langle V_{h_1} (z_1 ) V_{h_2} (z_2 ) W_{h_3} (z_3 ) \rangle \propto \frac{1}{|z_{12}|^{2 h_{12}-1}  |z_{13}|^{2 h_{13}+1}|z_{32}|^{2 h_{32}+1}}\, .
\label{VVW}
\end{align}
Later, we reproduce the $1/c$ corrections of these three point blocks from the networks of open Wilson lines.

Since the four point function of superfields depends on $(4|4)$ supercoordinates, it is parametrized by one bosonic and two fermionic superconformal invariants.
For them, we may choose (see, e.g., \cite{Belavin:2006zr,Belavin:2007eq})
\begin{align}
Z = \frac{Z_{12} Z_{34}}{Z_{42} Z_{31}} \, , \qquad
\tau_1 = \theta_{423} \, , \qquad \tau_2 = Z^{-1/2} (1-Z)^{-1/2} \theta_{421} \, .
\end{align}
For spinless operators, the four point function is then expressed as
\begin{align}
\left \langle \prod_{n=1}^4 \Phi_{h_n} (z_n , \theta_n)  \right \rangle &= 
|Z_{13}|^{- 4 h_1}  |Z_{24}|^{2 (h_3 - h_1 - h_2 - h_4)}|Z_{23}|^{2 (h_1 + h_4 - h_2 - h_3)} |Z_{34}|^{2(h_1 +h_2 - h_3 - h_4)} \nonumber \\
& \times \left[ G_0 (z , \bar z) + G_1 (z , \bar z) \bar \tau_1 \tau_1 
+  G_2 (z , \bar z) \bar \tau_2 \tau_2 + G_3 (z , \bar z)  \bar \tau_1 \tau_1  \bar \tau_2 \tau_2  \right] \, . \label{ssSCB}
\end{align}
This, in particular, means that there are four types of  independent superconformal blocks, such as, $\langle VVVV \rangle$, $\langle VVVW \rangle$, $\langle WVVV \rangle$, and $\langle WVWV \rangle$. The other types are obtained simply by the action of superconformal transformations. In this paper, we consider the $1/c$ corrections in the four point function $\langle VVVV \rangle$, but the extension to the other cases should be straightforward.

We compute superconformal blocks for the four point function from the bulk viewpoints.
We denote the operator product expansions schematically as
\begin{align}
V_{h_1} (z) V_{h_2} (0)= \sum_{p} z^{h_p - h_1 - h_2} \left( C_{12}^p [V_{h_p} (0)] _\text{e} + \tilde C_{12}^p [V_{h_p} (0)]_\text{o} \right) \, .
\label{ope}
\end{align}
We have again expressed the holomorphic sector only.
The superconformal families of $V_{h}$ are denoted as $[V_h]_\text{e}$ and $[V_h]_\text{o}$, which include descendants with integer and half-integer levels, respectively.
Following \cite{Zamolodchikov:1990ww,Belavin:2006zr,Belavin:2007eq},  we denote the superconformal families as
\begin{align}
[V_h (0)]_\text{e,o} = \sum_{N } z^N | N \rangle _{12}\, , \qquad |N \rangle  _{12} = C^{12}(N,h) V_h \, ,
\end{align}
where the sums are over $N = 0,1,2 \ldots $ for $[V_h]_\text{e}$ and
$N = 1/2,3/2,5/2 \ldots $ for $[V_h]_\text{o}$.
Here the vectors $|N \rangle _{12}$ are generated by acting operators $C^{12}(N,h)$ 
(see \eqref{CNh0} below for their expressions at the large $c$ limit)  to $V_h$.
With \eqref{ope}, the four point function can be decomposed in terms of superconformal blocks as
\begin{align}
\begin{aligned}
&\langle V_{h_1} (z) V_{h_2} (0) V_{h_3} (\infty) V_{h_4} (1) \rangle  \\
& \qquad = \sum_{p} C^p_{12} C^{p}_{34} |F_\text{e}  (h_i; h_p ;z)|^2 +  
\sum_{p} \tilde C^p_{12} \tilde C^{p}_{34} |F_\text{o} (h_i ; h_p ; z)|^2 \, ,
\end{aligned}
\label{CBdec}
\end{align}
where
\begin{align}
\begin{aligned}
&F_\text{e}  (h_i; h_p ;z) = z^{h - h_1 - h_2} \sum_{n=0}^\infty z^n  {}_{12} \langle n | n \rangle_{43} \,  , \\
&F_\text{o}  (h_i; h_p ;z) = z^{h - h_1 - h_2} \sum_{n=0}^\infty z^n  {}_{12}\langle n +\tfrac12 | n + \tfrac12 \rangle _{43} \, .
\end{aligned}
\end{align}
In order to evaluate them, we need the forms of the operators $C^{12}(N,h)$ explicitly.
They can be computed purely from the superconformal algebra but they are quite complicated in general. 
At the large $c$ limit, however, the operators $C^{12}(N,h)$ become simplified as (see (4.23) of  \cite{Belavin:2007eq})
\begin{align}
\begin{aligned}
&C^{12}(n,h) = \frac{(h+h_1 - h_2)_n}{n! (2 h)_{n}} G^{2n}_{-1/2} \, , \\
&C^{12}(n+\tfrac12,h) = \frac{(h+\tfrac12+ h_1 - h_2)_{n}}{n! (2 h)_{n+1}} G^{2n +1}_{-1/2} \, .
\end{aligned}
\label{CNh0}
\end{align}
Using
\begin{align}
G^{2n}_{1/2} G^{2n}_{-1/2} V_h = (2h)_n n! V_h \, , \quad
G^{2n+1}_{1/2} G^{2n+1}_{-1/2} V_h = (2h)_{n+1} n! V_h \, ,
\label{GGV}
\end{align}
the large $c$ limit of superconformal blocks can be computed as (see, e.g., \cite{Belavin:2006zr,Belavin:2007eq})
\begin{align}
\label{global0}
&F_\text{e} (h_i;h_p;z)  = z^{h_p - h_1 - h_2} {}_2 F_1 (h_p + h_1 - h_2 , h_p + h_4 - h_3 ; 2h_p ;z) + \mathcal{O}(c^{-1})\, ,  \\
&F_\text{o} (h_i;h_p;z) = \frac{z^{h_p - h_1 - h_2+1/2}}{2 h}    {}_2 F_1 (h_p + h_1 - h_2 + \tfrac12, h_p + h_4 - h_3 + \tfrac12 ; 2h_p +1 ;z) + \mathcal{O}(c^{-1})\, .
\nonumber
\end{align}
With general $c$, we can obtain superconformal blocks by solving differential equations \cite{Belavin:2006zr,Belavin:2007gz} in case that degenerate operators are involved.
It is difficult to obtain their closed forms but not so to find out their expressions in $z$ expansion, see appendix \ref{CFT} for the details. Another way is to use recursion relations as in \cite{Belavin:2006zr}, and the first few terms in $z$ expansion were obtained. 
The recursion relations are generalizations of the bosonic counter parts obtained in \cite{Zamolodchikov:1985ie}, see also \cite{Hadasz:2006qb,Hadasz:2008dt} for other types of recursion relations for the $\mathcal{N}=1$ superconformal blocks.

\section{Conformal weight of primary operator}
\label{Wilson}

In this section, we evaluate a two point function from an open Wilson line in the $\text{osp}(1|2)$ Chern-Simons gauge theory. In the next subsection, we give our strategy of Wilson line computation by focusing on how to renormalize divergences from loop diagrams developed in \cite{Hikida:2017ehf,Hikida:2018dxe}.
In subsection \ref{2pt1/c2}, we reproduce the conformal weight of primary operator up to $1/c^2$ order in \eqref{hexp} with the bulk method.

\subsection{Strategy of Wilson line computation}

We examine the 2d $\mathcal{N}=1$ super Virasoro minimal model with large $c$ in terms of the $\text{osp}(1|2) \oplus \text{osp}(1|2)$ Chern-Simons gauge theory, which describes a $\mathcal{N}=1$ supergravity theory on AdS$_3$ \cite{Achucarro:1987vz}.
The action is given by
\begin{align}
S = S_\text{CS} [A] - S_\text{CS} [\tilde A] \, , \quad
S_\text{CS}[A] = \frac{k_\text{CS}}{4 \pi} \int \text{tr} \left( A \wedge d A + \frac{2}{3} A \wedge A \wedge A \right) \, .
\label{CSaction}
\end{align}
The gauge fields $A, \tilde A$ take values in the osp$(1|2)$ Lie superalgebra, see, e.g., \cite{Frappat:1996pb} for some basics of Lie superalgebra.
We use the generators of the Lie superalgebra as $V^{2}_n$, $V^{3/2}_r$ with $n=\pm 1 ,0$, $r = \pm 1/2$. The generators $V^{2}_n$, $V^{3/2}_r$ satisfy
\begin{align}
\begin{aligned}
&[V^{2}_m , V^{2}_n] = (m-n) V^2_{m+n}  \, , \quad
[V^2_m , V^{3/2}_r] =  \left( \frac{m}{2} - r \right) V^{3/2}_{m+r} \, , \\
&\{ V^{3/2}_r , V^{3/2}_u \} = 2 V^2_{r + u}  \delta_{r+u}\, .
\end{aligned}
\end{align}
In a gauge, solutions to the equations of motion can be written as%
\footnote{In this paper, we consider  only the gauge field $A$.}
\begin{align}
A = e^{- \rho V_0^2} a(z) e^{\rho V_0^2} d z + V_0^2 d \rho \, .
\end{align} 
The AdS solution can be given by $a(z) = V_1^2$, which corresponds to the AdS metric
$ds^2 = d \rho^2 + e^{2 \rho} dz d \bar z$. Assigning that the configuration approaches to the AdS one for $\rho \to \infty$, we can set \cite{Banados:1998pi,Henneaux:1999ib} (see \cite{Campoleoni:2010zq,Henneaux:2010xg} for higher spin gravity and \cite{Hikida:2012eu,Henneaux:2012ny,Hanaki:2012yf} for higher spin supergravity)
\begin{align}
a (z) = V_1^2 (z) + \frac{1}{k_\text{CS}} \left( T(z) V_{-1}^2 + \frac12 G(z )  V^{3/2}_{-1/2} \right) \, .
\label{az}
\end{align}
It was shown that $T(z)$ and $G(z)$ generate the $\mathcal{N}=1$ superconformal algebra with the central charge
\begin{align}
c = 6 k_\text{CS} \, .
\end{align}
This value is the same as the well known Brown-Henneaux one \cite{Brown:1986nw}.

As in \cite{Besken:2017fsj,Hikida:2017ehf,Hikida:2018dxe},
we compute the two point function $\langle VV \rangle $  from  
\begin{align}
W_{(1)j} (z_f ; z_i)  \equiv \langle \text{lw} | W(z_f ; z_i) | \text{hw} \rangle \, .
\label{lwhw}
\end{align}
Here the open Wilson line is defined as
\begin{align}
W(z_f;z_i) = P \exp \left( \int_{z_i}^{z_f} dz  \, a(z ) \right) \, ,
\label{Wilsonop}
\end{align}
where $P$ denotes the path ordering and $\rho$-dependence is removed by a gauge transformation.
The finite dimensional representation of osp$(1|2)$ is labeled by $j=0,1/2,1,\cdots$ (see, e.g.,  \cite{Frappat:1996pb}).
In the above expression, $| \text{hw} \rangle$ and $|\text{lw} \rangle$ denote the highest and lowest weight states of the representation $j$, respectively.%
\footnote{The bulk symmetry is $\text{osp}(1|2) \oplus \text{osp}(1|2)$, which is enhanced to the sum of two $\mathcal{N}=1$ super Virasoro algebras near the boundary. From this, we can identify the global part of the super Virasoro algebra as the bulk osp$(1|2)$. The condition of the super Virasoro primary is given by \eqref{Primary}, and the global part of the condition leads to the highest weight state of the bulk osp$(1|2)$. Similar argument can be applied to the lowest weight state of the bulk osp$(1|2)$.}
We also use 
\begin{align}
W_{(2)j} (z_f ; z_i)  \equiv 
\langle \widehat{\text{lw}} | W (z_f ; z_i) | \widehat{\text{hw}} \rangle \, , \qquad
| \widehat{\text{hw}} \rangle  = V^{3/2}_{1/2} | \text{hw} \rangle  \, , \qquad
| \widehat{\text{lw}} \rangle  = V^{3/2}_{-1/2} | \text{lw} \rangle
\label{hlwhhw}
\end{align}
as a bulk computation for $\langle \Lambda \Lambda \rangle $.

We evaluate the path integral for the expectation value of the open Wilson line by integrating over the gauge fields. As explained in \cite{Fitzpatrick:2016mtp,Besken:2017fsj}, it can be effectively done by computing the products of $\mathcal{N}=1$ superconformal currents $T(z),G(z)$ inside the Wilson line operator \eqref{Wilsonop} with \eqref{az} in terms of correlators of a $\mathcal{N}=1$ superconformal theory.
We should integrate over the position $z$ of $T(z),G(z)$ inside the Wilson line operator \eqref{Wilsonop} with \eqref{az}, and there would be divergences arises from the coincident points of at least two of $T(z),G(z)$. Following \cite{Hikida:2017ehf,Hikida:2018dxe}, we regularize divergences by shifting the conformal weights of $T(z),G(z)$ as $2 \to 2 - \epsilon, 3/2 \to 3/2 - \epsilon$. Namely, we use the correlators of $T(z),G(z)$, such as,
\begin{align}
\langle T(z_2) T(z_1) \rangle = \frac{c/2}{z_{21}^{4-2\epsilon}} \, , \quad
\langle G(z_2) G(z_1) \rangle = \frac{2c/3}{z_{21}^{3-2\epsilon}} \, , \label{reg}
\end{align}
see \eqref{regulator} for others. 
In order to remove terms diverging at $\epsilon \to 0$, we introduce parameters $c_s$ $(s=2,3/2)$ with
\begin{align}
c_s = 1 + \frac{1}{c} c_s^{(1)} + \mathcal{O} (c^{-2}) 
\label{csexp}
\end{align}
and insert them in the Wilson line operator as
\begin{align}
W(z_f;z_i) = P \exp \left \{ \int_{z_i}^{z_f} dz \left[ V_1^2 (z) + \frac{6}{c} \left(c_2 T(z) V_{-1}^2 + \frac{c_{3/2}}{2} G(z )  V^{3/2}_{-1/2} \right) \right] \right\} \, .
\label{Wilsonopr}
\end{align}
We claim that  all divergences can be removed by renormalizing the parameters $c_s$ along with the normalization of the Wilson line such as
\begin{align}
Z_V^2 \langle W_{(1) j} (z) \rangle = \frac{1}{z^{2 h_j}}\, , \qquad
Z_W^2 \langle W_{(2)j} (z) \rangle = - \frac{2h_j}{z^{2 h_j +1}}\, . 
\label{ZVW}
\end{align}
Notice that there is no explicit relation between $Z_V$ and $Z_W$
since the introduction of the regulator $\epsilon$ in \eqref{reg} breaks the supersymmetry.

We choose $Z_{V}$, $Z_W$, and $c_{s}$  $(s = 2,3/2)$ such that there are no divergences at the limit $\epsilon \to 0$. However, there are terms surviving at the limit $\epsilon \to 0$, and we have to decide how to deal with them. Since we can fix the $\epsilon$-independent terms in $Z_{V}$ and $Z_W$ from \eqref{ZVW}, the problem is for $c_s$. In fact, the authors in \cite{Besken:2017fsj} failed to reproduce the $1/c^2$ order corrections in the conformal weight of primary operators in the Virasoro minimal model due to this issue. In \cite{Hikida:2017ehf,Hikida:2018dxe}, we have offered a prescription to fix $\epsilon$-independent terms in $c_s$ by requiring that the boundary correlators are consistent with the boundary symmetry. Here we adopt the same prescription. Concretely, we fix the $\epsilon$-independent terms in $c_2$ and $c_{3/2}$ such as to be consistent with the $\mathcal{N}=1$ superconformal Ward-Takahashi identities
\begin{align}
\lim_{y \to - \infty} |y|^{4} \langle V_h (z) V_h (0) T(y) \rangle = \frac{h z^2}{z^{2h}} \, , \quad
\lim_{y \to - \infty} |y|^{3} \langle \Lambda_h  (z) V_h (0) G(y) \rangle = - \frac{2 h z}{z^{2h}} \, ,
\label{WT}
\end{align}
see appendix \ref{N1WT} for a derivation of the right equation.
In \cite{Hikida:2017ehf}, we have solved the problem on $1/c^2$ order corrections in \cite{Besken:2017fsj} and extended to the case with the W$_3$ minimal model.

In the analysis of sl(2), it was convenient to work with the $x$-basis as in \cite{Verlinde:1989ua,Fitzpatrick:2016mtp,Hikida:2017ehf,Hikida:2018dxe}. In the current case with osp$(1|2)$, we introduce a bosonic parameter $x$ and a fermionic one $\xi$ as 
\begin{align}
\int dx d \xi 
\langle \text{lw} | x, \xi \rangle W_j (z_f ; z_i) \langle x , \xi | \text{hw} \rangle \, , \quad
\int dx d \xi 
\langle \widehat{\text{lw}} | x, \xi \rangle W_j (z_f ; z_i) \langle x , \xi | \widehat{\text{hw}} \rangle \, .
\end{align}
In terms of $ x,\xi$, the osp$(1|2)$ generators are represented as
\begin{align}
\begin{aligned}
&V^{2}_1 = \partial_x \, , \quad V^2_0 = x \partial_x - j + \frac12 \xi \partial_\xi \, , \quad V^2_{-1} = x^2 \partial_x - 2 j x + x \xi \partial_\xi \, , \\
&V^{3/2}_{1/2} = \xi \partial_x + \partial_\xi \, , \quad
V^{3/2}_{-1/2} = (x \partial_{x} - 2 j) \xi + x \partial_\xi \, .
\end{aligned}
\label{generators}
\end{align}
The lowest weight state is $ \langle x , \xi | \text{lw} \rangle  = 1$, and the next one is proportional to  $V^{3/2}_{-1/2} \langle x , \xi | \text{lw} \rangle = - 2 j \xi$.
Thus the dual states are
\begin{align}
\langle \text{lw} | x, \xi \rangle  = \xi \delta(x) \, , \quad
\langle \widehat{\text{lw}} | x, \xi \rangle  =  - \delta(x) \, ,
\label{lw}
\end{align}
where the normalizations are chosen for convenience.
The highest weight state is determined from the condition that it is annihilated by the actions of $V^{2}_{-1}$ and $V^{3/2}_{-1/2}$. Thus we find
\begin{align}
\langle x , \xi | \text{hw} \rangle  = x^{2j} \, , \quad
\langle x , \xi | \widehat{\text{hw}} \rangle  = - V^{3/2}_{1/2} \langle x , \xi | \text{hw} \rangle  =  - 2j  \xi  x^{2j - 1} 
\end{align}
with our choice of normalizations.

With these expressions, the open Wilson line can be expanded in $1/c$ as
\begin{align}
\begin{aligned}
W_{(a) j} (z_f;z_i)  =&  \sum_{n=0}^\infty \left( - \frac{6}{c}\right)^n \int_{z_i}^{z_f} dz_n \cdots \int_{z_i}^{z_2} dz_1  
\\ &  \times \sum_{s_j =2,3/2} \left [ \prod_{j=1}^n \frac{c_{s_j}}{N_{s_j}} J^{(s_j)} (z_j) \right ] f_{(a) j}^{(s_n , \ldots , s_1)} (z_n ,\ldots ,z_1) \, 
\end{aligned}
\label{Waj}
\end{align}
with
\begin{align}
J^{(2)}(z) = T(z) \, , \quad J^{(3/2)}(z) = G(z) \, , \quad
N_2 = -1 \, , \quad N_{3/2} = - 2 \, . 
\label{Ns}
\end{align}
Moreover, we defined
\begin{align}
&f_{(1)j}^{(s_n , \ldots , s_1)} (z_n ,\ldots ,z_1) 
= \int d \xi \xi \left. V^{s_n} (z_{fn}) \cdots V^{s_1} (z_{f1}) (x + z_f - z_i)^{2j} 
\right|_{x = 0} \, , \\
&f_{(2)j}^{(s_n , \ldots , s_1)} (z_n ,\ldots ,z_1) 
= \int d \xi  \left. V^{s_n} (z_{fn}) \cdots V^{s_1} (z_{f1}) 2j \xi (x + z_f - z_i)^{2j-1} 
\right|_{x = 0} 
\end{align}
with $z_{fj} \equiv z_f - z_j$ and
\begin{align}
V^{s} (z) \equiv   e^{z V_1^2} V^{s}_{-s+1} e^{- z V_1^2} \, .
\label{Vs}
\end{align}

\subsection{Two point function up to $1/c^2$ order}
\label{2pt1/c2}

In this subsection, we evaluate the expectation value of the open Wilson line $ \langle W_{(1)j} (z;0) \rangle$ up to $1/c^2$ order. 
We first evaluate $ \langle W_{(1)j} (z;0) \rangle$ up to $1/c$ order. We reproduce the conformal weight \eqref{hexp} and fix $Z_V$ up to this order. We also examine  $ \langle W_{(2)j} (z;0) \rangle$ since we need the expression of $Z_W$ to determine $c_{3/2}$. We then compute the three point function with $T(z)$ or $G(z)$ and fix $c_2$ and $c_{3/2}$ through the superconformal Ward-Takahashi identities \eqref{WT}. With the information of $Z_V$, $c_2$, and $c_{3/2}$ up to the order $1/c$, we reproduce  the conformal weight \eqref{hexp}  up to the $1/c^2$ order from the open Wilson line $ \langle W_{(1)j} (z;0) \rangle$.

We start from the expectation value of the open Wilson line $ \langle W_{(1)j} (z;0) \rangle$.
At the leading order in $1/c$, we find
\begin{align}
\langle W_{(1)j}(z;0) \rangle = z^{2j} + \mathcal{O}(c^{-1})
\end{align}
as expected. At the next leading order, there are two contributions 
\begin{align}
I^{(1)}_{s} (z)= \left(- \frac{6}{c} \right)^2 \frac{1}{N_s^2} \int_0^z dz_2 \int_0^{z_2} dz_1 f^{(s,s)}_{(1)j} (z;0;z_2 ,z_1) \langle J^{(s)} (z_2) J^{(s)} (z_1) \rangle  
\label{I2s}
\end{align}
with $s=2,3/2$.  Here $J^{(s)}$ and $N_s$ were introduced in \eqref{Ns}. Moreover, we have used $c_s = 1 + \mathcal{O}(c^{-1})$.
Introducing the regulator $\epsilon$ as in \eqref{reg}, 
the sum of them is computed as
\begin{align}
I^{(1)}_{2} (z)+ I^{(1)}_{3/2}(z) = \frac{z^{2j}}{c} \left[ \frac{3 j (2 j+1) }{\epsilon}+j  (6 (2 j+1) \log (z)+10 j+1) \right] + \mathcal{O} ( \epsilon) \, .
\end{align}
Therefore, by setting
\begin{align}
Z_V^2  = 1 - \frac{1}{c}\left( \frac{3j(2j+1)}{\epsilon} + j (10j+1) \right) \, ,
\label{ZV}
\end{align}
we reproduce \eqref{ZVW} with the correct shift of conformal weight as in \eqref{VVexp} with \eqref{hexp}.
In the same manner, we have
\begin{align}
\begin{aligned}
&\langle W_{(2)j}(z;0) \rangle = 2 j z^{2j -1} 
\\& \quad + \frac{2 j z^{2j -1}}{c} \left[ \frac{3 j (2 j+1) }{ \epsilon } + 6 j (2 j+1) \log (z)+j (10 j+3)+2 \right]  + \mathcal{O}(c^{-2})
\end{aligned}
\end{align}
up to the term vanishing at $\epsilon \to 0$.
Thus the normalization
\begin{align}
Z_W^2 = 1 - \frac{1}{c}\left( \frac{3 j (2 j+1) }{ \epsilon } + 10 j^2 - 3 j - 1 \right) 
\label{ZW}
\end{align}
leads to \eqref{ZVW} with the same shift of conformal weight.

In order to fix $c_2$ and $c_{3/2}$, we need to evaluate the three point function with $T(y)$ or $G(y)$ as in \eqref{WT} with an open Wilson line.   
For the three point function with $T(y)$, we compute
\begin{align}
H^{(2)} (z) = \lim_{y \rightarrow - \infty} |y|^{4} \langle W_{(1)j}(z;0) T(y) \rangle 
\label{3pt}
\end{align}
as argued in \cite{Hikida:2017ehf,Hikida:2018dxe}.
At the leading order in $1/c$, this becomes
\begin{align}
H^{(2)} (z)|_{\mathcal{O}(c^0)} =  \left. -\frac{6}{c} \frac{c_2}{N_2}\int^z_0dz_1 f^{(2)} _{(1)j} (z_1) \lim_{y \rightarrow - \infty} |y|^{4 } \langle T(z_1) T(y) \rangle \right |_{\mathcal{O}(c^0)} 
=
-jz^{2j+2} \, ,
\end{align}
which is consistent with \eqref{WT}.
At the next leading order in $1/c$, there are three types of contribution.
The first one comes from the above integral with the $1/c$ order term in $c_2 = 1 + c_2^{(1)} /c + \mathcal{O} (c^{-2})$ as 
\begin{align}
H^{(2)}_{2} (z)= - \frac{j c_2^{(1)}}{c} z^{2j+2} \, .
\end{align}
The second type is with the insertions of two extra currents as
\begin{align}
\begin{aligned}
H^{(2)}_{ss} (z) =& \left(-\frac{6}{c}\right)^2 \frac{1}{N_s^2 } \int^z_0 dz_2\int^{z_2}_0 dz_1 f^{(s,s)} _{(1)j} (z_2, z_1)  \\ &\times \lim_{y \rightarrow - \infty} |y|^{4 - 2 \epsilon} \langle J^{(s)}(z_2) J^{(s)}(z_1) T(y) \rangle 
\end{aligned}
\end{align}
with $s=2,3/2$. The third type is with the insertions of three extra currents as
\begin{align}
\begin{aligned}
H^{(2)}_{s_3 s_2 s_1} (z)=& \left(-\frac{6}{c}\right)^3 \frac{1}{N_{s_3} N_{s_2} N_{s_1}} \int^z_0 dz_3\cdots\int^{z_2}_0dz_1 f^{(s_3, s_2,s_1)} _{(1)j}(z_3,z_2,z_1) \\ &\times
\lim_{y\rightarrow - \infty} |y|^{4 - 2 \epsilon} \langle J^{(s_3)}(z_3) J^{(s_2)}(z_2) J^{(s_1)}(z_1) T(y) \rangle \, .
\end{aligned}
\end{align}
Non-trivial contributions are with 
\begin{align}
(s_1,s_2,s_3) = (2,2,2), (\tfrac32,\tfrac32,2),(\tfrac32,2,\tfrac32),(2,\tfrac32,\tfrac32) \, .
\end{align}
For the second and third types, we use the correlators of currents
in \eqref{regulator}.
The sum of them can be written as
\begin{align}
Z_V^2 H^{(2)} (z) = \left[-j +\frac{j}{c} \left( \frac{9/2}{\epsilon} - 6j -1 - c_2^{(1)}\right) \right] \frac{z^{2}}{z^{2 h_j}} 
\end{align}
up to the order $1/c$. The factor $Z_V^2$ obtained in \eqref{ZV} is included such that the operator $V_h (z)$ in the three point function \eqref{WT} is normalized as in \eqref{ZVW}.
The relation in \eqref{WT} is realized if we choose the parameter $c_2$ as
\begin{align}
c_2=1+\frac{1}{c}\left(\frac{9/2}{\epsilon}+2\right) + \mathcal{O}(c^{-2}) \, .
\label{c2r}
\end{align}
In this way, we fix the $1/c$ order term in $c_2$ along with removing the divergence from the $1/\epsilon$ order term in $\epsilon \to 0$.

Next, we evaluate the three point function with $G(y)$ in \eqref{WT} using an open Wilson line. 
For this, we define
\begin{align}
W_{(3)j} (z_f ; z_i) \equiv \langle \widehat{\text{lw}} | W (z_f ; z_i) | \text{hw} \rangle \, ,
\label{W3j}
\end{align}
which can be expanded in $1/c$ as in \eqref{Waj} with
\begin{align}
f_{(3)j}^{(s_n , \ldots , s_1)} (z_n ,\ldots ,z_1) 
= - \int d \xi  \left. V^{s_n} (z_{fn}) \cdots V^{s_1} (z_{f1}) (x + z_f - z_i)^{2j} 
\right|_{x = 0} \, .
\end{align}
Then, the three point function is evaluated from 
\begin{align}
Z_W Z_V  H^{(3/2)} (z) = \lim_{y \rightarrow - \infty} |y|^{3} Z_W Z_V  \langle W_{(3)j}(z;0) G(y) \rangle \, .
\label{3ptf}
\end{align}
Here $Z_W$  and  $Z_V$ obtained in \eqref{ZW} and \eqref{ZV}  are included such that $\Lambda _h (z)$ and $V_h (0)$ in the three point function are normalized as in \eqref{ZVW}.
At the leading order in $1/c$, we have
\begin{align}
\begin{aligned}
H^{(3/2)} (z)|_{\mathcal{O}(c^0)} &=  \left. -\frac{6}{c} \frac{c_{3/2}}{N_{3/2}} \int^z_0dz_1 f^{(3/2)}_{(3)j}(z_1) \lim_{y\rightarrow\infty} |y|^{3} \langle G(z_1) G(y) \rangle \right|_{\mathcal{O}(c^0)}\\
&=  2 j z^{2 j +1} 
\end{aligned}
\label{H321}
\end{align}
as in \eqref{WT}.

Just as in the case with $H^{(2)} (z)$, there are three types of contribution at the next leading order.
The first one comes from \eqref{H321} but with $c_{3/2}^{(1)}$ in $c_{3/2} = 1 + c^{(1)}_{3/2}/c + \mathcal{O}(c^{-1})$ as
\begin{align}
H^{(3/2)}_{3/2} (z)=  \frac{2 j c_{3/2}^{(1)}}{c} z^{2 j +1} \, .
\end{align}
The second one are with the insertion of two extra currents as
\begin{align}
\begin{aligned}
H^{(3/2)}_{s s'} (z)=&  \left( - \frac{6}{c} \right)^2 \frac{1}{N_2 N_{3/2} }\int^z_0 dz_2\int^{z_2}_0 dz_1 f^{(s,s ')}_{(3)j}(z_2,z_1) \\
&\times \lim_{y \rightarrow - \infty} |y|^{3 - 2 \epsilon} \langle J^{(s)}(z_2) J^{(s')}(z_1) G(y) \rangle
\end{aligned}
\end{align}
with $(s , s') = (2,3/2),(3/2 ,2)$.
The third one are with the insertion of three extra currents as
\begin{align}
\begin{aligned}
H^{(3/2)}_{s_3 s_2 s_1} (z)=& \left( - \frac{6}{c} \right)^3 \frac{1}{N_{s_3}N_{s_2}N_{s_1}}\int^z_0 dz_3 \int^{z_3}_0dz_2  \int^{z_2}_0dz_1 f^{(s_3 ,s_2 , s_1)}_{(3)j}(z_3,z_2,z_1)\\
&\times \lim_{y \rightarrow - \infty} |y|^{3- 2 \epsilon} \langle J^{(s_3)}(z_3)J^{(s_2)}(z_2)J^{(s_1)}(z_1) G(y) \rangle \, .
\end{aligned}
\end{align}
The non-trivial contributions are with
\begin{align}
(s_1 ,s_2 , s_3) = (2,2,\tfrac32) , (2, \tfrac32 ,2) , (\tfrac32 ,2,2), (\tfrac32,\tfrac32,\tfrac32) \, .
\end{align}
From the sum of these contributions, we find
\begin{align}
\begin{aligned}
Z_W Z_V H^{(3/2)} (z) = \left[ 2j - \frac{j}{c}\left(\frac{9}{\epsilon} - 12 j - 3 -2c_{3/2}^{(1)}\right) \right] \frac{z}{z^{2 h_j}} 
\end{aligned}
\end{align}
up to  $1/c$ order.
Choosing the parameter $c_{3/2}$ as
\begin{align}
c_{3/2}=1+\frac{1}{c}\left(\frac{9/2}{\epsilon}+\frac{3}{2}\right) + \mathcal{O}(c^{-2}) \, ,
\label{c32r}
\end{align}
we reproduce \eqref{WT} up to the first non-trivial order in $1/c$.

Now that we have $Z_V$, $c_2$, and $c_{3/2}$ as in \eqref{ZV}, \eqref{c2r}, and \eqref{c32r}, we can read off the $1/c^2$ order term in \eqref{hexp} from the expectation value of the open Wilson line
\begin{align}
\left. Z_V^2 \langle W_{(1)j} (z ;0) \rangle \right|_{\mathcal{O}(c^{-2})}
\end{align} 
as in \eqref{VVexp}.
There are four types of contribution to this quantity.
We have already evaluated the $1/c$ order term in $ \langle W_{(1)j} (z ;0) \rangle $ above as the sum of $I_2^{(2)}(z)$ and $I_2^{(3/2)}(z)$ defined in \eqref{I2s}. Therefore, the first contribution is with the $1/c$ order term in $Z_V^2$ as
\begin{align}
I^{(2)}_Z (z)= - \frac{1}{c}\left( \frac{3j(2j+1)}{\epsilon} + j (10j+1) \right) (I^{(1)}_2(z) + I^{(1)}_{3/2}(z)) \, .
\end{align}
For the computation of $ \langle W_{(1)j} (z ;0) \rangle $  at the order of $1/c$, we have set $c_2 = c_{3/2} =1$. Therefore, the second type of contribution is
\begin{align}
I_c^{(2)} (z) = \frac{ 2 c_2^{(1)}}{c} I^{(1)}_2(z)+ \frac{2 c_{3/2}^{(1)}}{c} I^{(1)}_{3/2}(z)  = 
\frac{1}{c}\left( \frac{9}{ \epsilon} + 4 \right) I^{(1)}_2(z) + \frac{1}{c}\left( \frac{9}{ \epsilon} + 3  \right) I^{(1)}_{3/2}(z) \, .
\end{align}
The third type are with the insertion of three currents as
\begin{align}
\begin{aligned}
I^{(2)}_{s_3 s_2 s_1} (z)=&\left(-\frac{6}{c}\right)^3 \frac{1}{N_{s_3} N_{s_2} N_{s_1}} \int^{z}_0 dz_3 \int^{z_3}_0 dz_3 \int^{z_2}_0 dz_1 \\
& \times f^{(s_3,s_2,s_1)}_{(1)j}(z_3, z_2, z_1) 
\langle J^{(s_3)}(z_3) J^{(s_2)}(z_2) J^{(s_1)}(z_1) \rangle 
\end{aligned}
\end{align}
with
\begin{align}
(s_1 ,s_2 ,s_3) = (2,2,2) , (2 , \tfrac32 ,\tfrac32 ) , (\tfrac32 , 2, \tfrac32) , (\tfrac32,\tfrac32,2) \, .
\end{align}
The fourth type are with the insertion of four currents as
\begin{align}
\begin{aligned}
I^{(2)}_{s_4 s_3 s_2 s_1} (z)=&\left(-\frac{6}{c}\right)^4\frac{1}{N_{s_4} N_{s_3} N_{s_2}N_{s_1}}\int^z_0 dz_4\cdots\int^{z_2}_0 dz_1 \\
& \times f^{(s_4, s_3,s_2,s_1)}_{(1)j}(z_4, z_3, z_2, z_1) 
\langle J^{(s_4)}(z_4) J^{(s_3)}(z_3) J^{(s_2)}(z_2) J^{(s_1)}(z_1) \rangle 
\end{aligned}
\end{align}
with
\begin{align}
\begin{aligned}
(s_1 ,s_2 ,s_3, s_4) = &  (2,2,2,2) , (2 , 2, \tfrac32 ,\tfrac32 ) , ( \tfrac32 ,\tfrac32 ,2 , 2 ) ,(2 , \tfrac32 , \tfrac32 , 2)  ,(\tfrac32 , 2 , 2,\tfrac32) , \\
& (2 , \tfrac32 , 2 ,  \tfrac32 ),  (\tfrac32, 2 ,  \tfrac32, 2 ) ,  (\tfrac32 , \tfrac32 ,  \tfrac32 , \tfrac32 )  \, .
\end{aligned}
\end{align}
For the correlators of currents, we use \eqref{regulator}.

Let us first sum over the third and fourth types of contribution as
\begin{align}
\begin{aligned}
&I^{(2)}_\text{sum} (z)=\frac{z^{2j}}{c^2}\Biggl[\frac{ 18j  (j-1) (2 j+1) (2 j+3) \log (z)}{ \epsilon }\\
& \qquad + 3j\log (z) \left(6 (2 j+1) \left(4 j^2+2 j-3\right) \log (z)+(2 j-1) \left(40 j^2+44 j+1\right)\right) \Biggr]  \, ,
\end{aligned}
\end{align}
where we have kept only the term including $\log (z)$.
From this, we observe that there is non-local divergence from the term proportional to $\log (z) /\epsilon $. Moreover, the terms in front of $\log (z)$ and $\log ^2(z)$ do not reproduce \eqref{VVexp} with \eqref{hexp}. This corresponds to the problem in \cite{Besken:2017fsj} for the bosonic case. The problem was solved in \cite{Hikida:2017ehf} by including $Z_V$ and $c_2$ in the current terminology. Similarly we arrive at
\begin{align}
\begin{aligned}
\left. Z_V^2 
\langle W _{ (1)j}(z) \rangle \right|_{\mathcal{O}(c^{-2})} &=
I^{(2)}_Z (z) + I^{(2)}_c (z) +  I^{(2)}_{\text{sum}} (z)  \\
& = \frac{z^{2j}}{c^2} \left[ 18j^2(2 j+1)^2 \log^2 (z) + 45j (2 j+1) \log (z) \right]
\end{aligned}
\end{align}
after including the effects of $Z_V$, $c_2$, and $c_{3/2}$.
In the final expression, there is no non-local term and the $1/c^2$ order correction of the conformal weight in \eqref{hexp} is correctly reproduced.

\section{Superconformal blocks}
\label{superCB}

From the networks of open Wilson lines, we study more generic correlators.
As in \cite{Bhatta:2016hpz,Besken:2016ooo,Besken:2017fsj,Hikida:2018dxe}, we evaluate
\begin{align}
G_n (j_i | z_i ) \equiv \langle S | \prod_{i=1}^n W_{j_i} (z_0 ; z_i) | \text{hw} \rangle_i 
\label{Gn}
\end{align}
for $n$-point blocks of the type $\langle \prod_{i=1}^n V  \rangle$.
Here $S$ represents a singlet state in the product representation $j_1 \otimes \cdots \otimes j_n$. 
We can construct a singlet from the product representation in many ways in general.
There are also many conformal blocks for a correlator in general as well.
As explained in \cite{Besken:2016ooo,Hikida:2018dxe}, a different choice of $S$ leads to a different  conformal block.  The open Wilson lines are connected by the vertex $\langle S |$ at $z=z_0$, but the expectation value should not depend on $z_0$. 
We evaluate the $1/c$ corrections for the Wilson line networks by following the analysis in \cite{Hikida:2018dxe}.
When we consider correlators involving $W$ (or $\Lambda$), then we should replace $ | \text{hw} \rangle $ by $ | \widehat{\text{hw}} \rangle $.

As in \cite{Hikida:2018dxe}, it is convenient to  move to the conjugate form of \eqref{Gn} and to use the basis in terms of $X_i = (x_i , \xi_i)$ as
\begin{align}
G_n (j_i|z_i) 
= \left[ \prod_{i=1}^n \int d X_i \langle \text{lw} | X_i \rangle W_{j_i} (z_i ; z_0 ; X_i) \langle X_i | \right] | S \rangle \, .
\label{Gnd}
\end{align}
The singlet condition is now written as
\begin{align}
\sum_{i=1}^n V^s_m (X_i) \left [ \prod_{i=1}^n \langle X_i |  \right] | S \rangle  = 0
\label{singlet}
\end{align}
for all $s=2,3/2$ and $m= -s+1, \cdots , s-1$. Here the generators $V^s_m (X_i)$ are given in \eqref{generators} with $X_i = (x_i , \xi _i)$.

In the next subsection, we reproduce the three point blocks in \eqref{VVV} and \eqref{VVW} up to  $1/c$ order. In subsection \ref{identity} and subsection \ref{general}, we consider the four point function
\begin{align}
\langle V_{h_q} (z)  V_{h_q} (0) V_{h_j} (\infty) V_{h_j} (1) \rangle
\label{4pt0}
\end{align}
with $h_q$ and $h_j$ in \eqref{hexp}. 
We obtain the closed form expressions of the identity block up to $1/c^2$ order and those of the general blocks up to $1/c$ order.
We check that the results are consistent with those obtained from a $\mathcal{N}=1$ superconformal field theory.

\subsection{Three point blocks}
\label{3ptblocks}

We consider the three point blocks in \eqref{VVV} and \eqref{VVW}.
In the expression of \eqref{Gnd} with \eqref{lw}, we evaluate
\begin{align}
G_3^{(1)} (j_i | z) = \int d \xi_1 \xi_1  \int d \xi_2 \xi_2 \int d \xi_3 \xi_3  \left. \left [ \prod_{i=1}^3 W_{j_i} (z_i ; z_0 ; X_i)  \right] v_3^{(1)} (j_i | X_i) \right|_{x_i = 0} 
\label{G31}
\end{align}
for \eqref{VVV} and 
\begin{align}
G_3^{(2)} (j_i | z) = \int d \xi_1 \xi_1  \int d \xi_2 \xi_2 \int d \xi_3  \left. \left [ \prod_{i=1}^3 W_{j_i} (z_i ; z_0 ; X_i)   \right] v_3^{(2)} (j_i | X_i) \right|_{x_i = 0 } 
\label{G32}
\end{align}
for \eqref{VVW} up to  irrelevant phase factors.
Here we have set $(z_1 ,z_2 ,z_3) = (z , 0 , 1)$ and $z_0 = z_3 =1$. 
Moreover, we have used the two independent solutions to the singlet condition \eqref{singlet}, which are given by 
\begin{align}
v_3^{(1)} (j_i | X_i) = X_{12}^{j_{12}} X_{13}^{j_{13}} X_{32}^{j_{32}} \, , \quad
v_3^{(2)} (j_i | X_i) = \xi_{123} X_{12}^{j_{12}} X_{13}^{j_{13}} X_{32}^{j_{32}}
\label{v3}
\end{align}
with
\begin{align}
X_{ij} = x_i - x_j - \xi_i \xi_j \, , \quad 
\xi_{123} = (X_{12} X_{13} X_{32} )^{-1/2} (X_{23} \xi_1 + X_{31} \xi_2 + X_{12} \xi_3 + \xi_1 \xi_2 \xi_3) \, ,
\end{align}
and $j_{12} = j_1 + j_2 - j_3$ and so on.
Noticing that 
\begin{align}
\left.  W_{j_i} (z_i ; z_0 ; X_i) \right|_{\mathcal{O}(c^0)} = e^{(z_i - z_0) \partial_{x_i}} \, ,
\label{leadingW}
\end{align}
we can reproduce the leading order expressions of \eqref{VVV} and \eqref{VVW} from \eqref{G31} and \eqref{G32}, respectively.

\begin{figure}
	\centering
	\includegraphics[keepaspectratio, scale=1]
	{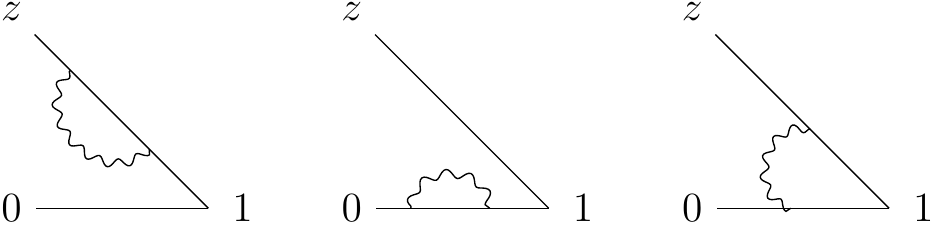}
	\caption{Contributions to the three point blocks from the networks of open Wilson lines at the $1/c$ order. The solid and wavy lines represent an open Wilson line and the propagation of a spin $s$ current with $s=2,3/2$.}
	\label{Wilson1}
\end{figure}

After these preparations, we move to the next leading order in $1/c$ and compute the contributions expressed by the diagrams of figure \ref{Wilson1}.
We define
\begin{align}
\begin{aligned}
&h^{s,\alpha} (z_b , z_a ; y_2 , y_1 ; x_3) \\
& \quad = \int d \xi_3 (\xi_3)^{2 - \alpha} \left. V^s (z_b - y_2) V^s (z_a - y_1) e^{(z-1) V^2_1 (x_1)} e^{ - V^2_1 (x_2)}  v_3^{(\alpha)} (j_i | X_i) \right|_{X_{1,2}=0} \, ,
\end{aligned}
\end{align}
where $V^s (z)$ is given in \eqref{Vs}.
The integrals to compute are 
\begin{align}
\begin{aligned}
&t_1^{s,\alpha} (z ; x_3 ) = \left( -\frac{6}{c }\right)^2 \frac{1}{N_s^2}\int_1^z dy_2 \int_1^{y_2} dy_1 h^{s,\alpha} (z , z ; y_2 , y_1 ; x_3) \langle J^{(s)} (y_2) J^{(s)} (y_1) \rangle \, , \\ 
&t_2^{s,\alpha} (z ; x_3 ) = \left( -\frac{6}{c }\right)^2 \frac{1}{N_s^2} \int_1^0 dy_2 \int_1^{y_2} dy_1 h^{s,\alpha}  (0 , 0 ; y_2 , y_1 ; x_3) \langle J^{(s)} (y_2) J^{(s)} (y_1) \rangle \, , \\ 
&t_3^{s,\alpha} (z ; x_3 ) = \left( -\frac{6}{c }\right)^2 \frac{1}{N_s^2} \int_1^0 dy_2 \int_1^{z} dy_1 h^{s,\alpha} (0 , z ; y_2 , y_1 ; x_3) \langle J^{(s)} (y_2) J^{(s)} (y_1) \rangle \, ,
\end{aligned}
\label{talpha}
\end{align}
where the two point functions of currents are given in \eqref{reg}.
The contributions at the $1/c$ order are given by the sums of these integrals as
\begin{align}
\left. \langle G_3^{(\alpha)} (j_i |z) \rangle \right|_{\mathcal{O}(c^{-1})} 
= \sum_{a=1}^3 \sum_{s=2,3/2} t_a^{s,\alpha} (z ; 0) \, ,
\end{align}
which are evaluated as
\begin{align}
\begin{aligned}
\left. \langle G_3^{(1)} (j_i |z) \rangle \right|_{\mathcal{O}(c^{-1})} 
& = -\frac{1}{c}z^{j_{12}} (z-1)^{j_{13}} ( h^{(1)}_{12} \log (z) + h^{(1)}_{13} \log (z -1 ) ) \, ,  \\
\left. \langle G_3^{(2)} (j_i |z) \rangle \right|_{\mathcal{O}(c^{-1})} 
& = -\frac{1}{c}z^{j_{12}+1/2} (z-1)^{j_{13}-1/2} ( h^{(1)}_{12} \log (z) + h^{(1)}_{13} \log (z -1 ) ) 
\end{aligned}
\end{align}
with
\begin{align}
h^{(1)}_i = - 3 j_i (2 j_1 + 1)  \, , \quad
h^{(1)}_{12} = h^{(1)}_1 + h^{(1)}_2 - h^{(1)}_3 
\end{align}
and so on. Here we have subtracted the terms proportional to the tree level expressions, which change only irrelevant overall constants.
We can see that they reproduce the $1/c$ order terms in \eqref{VVV} and \eqref{VVW}.

Before going into the analysis of four point blocks, we would like to remark on a point, which will be important later. There are two independent singlets $|S \rangle$  as in \eqref{v3}, and we chose one of them for \eqref{G31} and the other for \eqref{G32}. The choice sounds obvious from the analysis at the leading order in $1/c$, but it could be subtle at the higher orders.
An open Wilson line includes the information of both $V_h(z)$ and $\Lambda_h(z)$,
since it represents the propagation of an superfield.
An insertion of $G(z)$ exchange $V_h(z)$ and $\Lambda_h(z)$ as in \eqref{WT} and \eqref{W3j}, thus we should be careful on quantum effects associated with the insertions of $G(z)$. For three point blocks at $1/c$ order, this does not cause any problems.
This is because the number of $G(z)$ inserted should be even and the even number of $\xi_i$ action does not exchange the two solutions in \eqref{v3}. However, this point becomes important for four point blocks as seen below.

\subsection{Identity four point blocks}

\label{identity}

\begin{figure}
	\centering
	\includegraphics[keepaspectratio, scale=1]
	{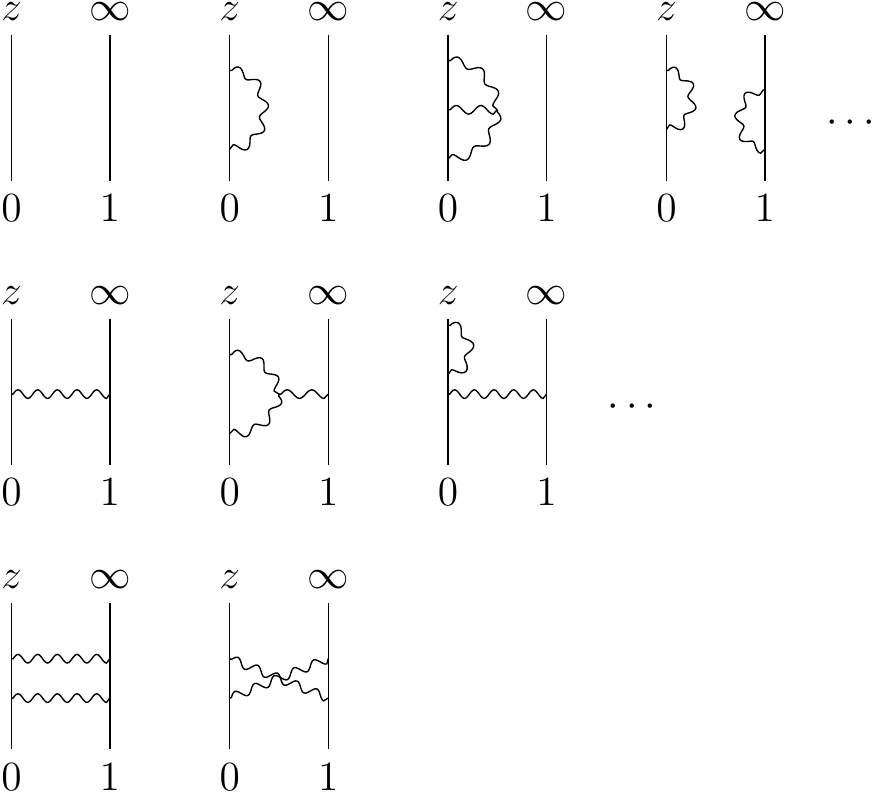}
	\caption{Several representative contributions to the identity four point block from the network of open Wilson lines up to the $1/c^2$ order. The top diagrams represent the contributions of the self-energy type. The middle diagrams represent the contribution with the exchange of energy momentum tensor and its $1/c$ corrections. The bottom diagrams represent the contributions with the exchange of two spin $s$ currents with $s=2$ or $s=3/2$.}
	\label{Wilson2}
\end{figure}

As was shown in \cite{Hikida:2018dxe}, the network of open Wilson lines for the identity four point block can be deformed to the product of two open Wilson lines as
\begin{align}
\langle W_q (z ;0) W_j (\infty ; 1) \rangle \, ,
\end{align}
see also \cite{Fitzpatrick:2016mtp}.
There are three types of contributions to the $1/c$ corrections as illustrated with the diagrams of figure  \ref{Wilson2}. 
See figure 1 and figure 2 in \cite{Hikida:2018dxe} for the diagrams omitted here.
The contributions coming from the top diagrams are of the self-energy type. We do not study the contributions here since we already knew that the net effect only shifts the conformal weight. 

The middle left diagram describes the exchange of a current. 
Noticing that only the energy momentum tensor (or an integer spin current) is exchanged, the contribution from the diagram is
\begin{align}
\begin{aligned}
K^{(1)} (z) &= \left(- \frac{6}{c}\right)^2 \frac{1}{N_2^2} \int_0^z d z_2 \int_1^\infty d z_1 f^{(2)}_{(1)q} (z;0;z_2) 
f^{(2)}_{(1)j} (\infty ; 1 ; z_1 ) \langle T(z_2) T(z_1) \rangle  \\
&= \frac{2 h_j  h_q}{c} z^{2 q + 2} {}_2 F_1 (2,2;4;z) 
\end{aligned}
\label{H1}
\end{align}
up to $1/c$ order, where $h_q = - q$ and $h_j = - j$.
With $q = 1/2$, the $z$ expansion of the above expression reproduces the $1/c$ order term in \eqref{iddeg}. 
The middle diagrams in figure \ref{Wilson2} describes a type of contributions, which  
can be included by using $h_q,h_j$ in \eqref{H1} with $1/c$ corrections as shown in \cite{Hikida:2018dxe}. Therefore, the $1/c^2$ order contributions of this type can be summarized as
\begin{align}
K^{(2)}_h (z) = \frac{12 \left(j^2 q+j q^2+j q\right) }{c^2}z^{2 q+2} \, _2F_1(2,2;4;z) \, .
\end{align}

Another type of contributions at the order of $1/c^2$ come from the exchanges of two currents as in
the bottom diagrams of figure \ref{Wilson1}.
In the current case, the spin (or more precisely the statistic) of two currents should be the same. Thus the contribution is the sum of $(s=2,3/2)$
\begin{align}
\begin{aligned}
K^{(2)}_{s}(z) = &\left(- \frac{6}{c} \right)^4 \frac{B_s^2}{N_s^4}  \int_0^z dz_4 \int_0^{z_4} d z_3 \int_1^\infty dz_2 \int_1^{z_2} dz_1 \\ &\times
f^{(s,s)}_q (z ; 0 ; z_4 ,z_3) f^{(s,s)}_j (\infty ; 1 ; z_2 ,z_1) \left( \frac{1}{(z_{41} z_{32})^{2 s}} +  \frac{(-1)^{2s}}{(z_{42} z_{31})^{2 s}}\right) 
\end{aligned}
\end{align}
with
\begin{align}
B_2 = \frac{c}{2}  \, , \quad
B_{3/2} =  \frac{2 c}{3} \, .
\end{align}
The sum  is computed as
\begin{align}
\begin{aligned}
&K^{(2)}_{2} (z) + K^{(2)}_{3/2} (z) 
= \frac{18 j q z^{2 q-2}}{c^2} \left(2 z^2 (8 j (q+1)+8 q+3) \right. \\ 
& \qquad +\log ^2(1-z) \left(4 j \left(q (z-2)^2+2 (z-1)\right)+(z-1) (8 q+3 z-4)\right) \\
& \qquad \left. -(z-2) z (4 j (4 q+1)+4 q-1) \log (1-z)+6 (z-2) z \text{Li}_2(z)\right)\, .
\end{aligned}
\end{align}
From the sum $\sum_{s=h,2,3/2} K_s^{(2)} (z)$, we obtain the identity block as
\begin{align}
\begin{aligned}
z^{2 h_q} \mathcal{V}_0 (z) =& 1 + \frac{2 q j }{c} z^2 {}_2 F_1 (2,2;4;z) \\
&+ \frac{1}{c^2} \left[ q^2 j^2 k_a (z) + (q^2 j +  q j^2) k_b (z) + q j k_c (z)\right]  + \mathcal{O} (c^{-3}) 
\end{aligned}
\end{align}
with
\begin{align}
&k_a (z) = 2 \left( z^2 {}_2 F_1 (2,2;4;z) \right)^2 \, , \nonumber \\
&k_b (z) = 144 \left( 1 + \frac{(z - 1 ) \log ^2(1-z)}{z^2} \right) \, , \\
&k_c(z) = \frac{18}{z^2} \left(6 (z-2) z \text{Li}_2(z)-2 z^2+(z-1) (3 z-4) \log ^2(1-z)+5 (z-2) z \log (1-z)\right) \, . \nonumber 
\end{align}
We can check that the result matches with the expression in \eqref{iddeg} by setting $q=1/2$.

\subsection{General four point blocks}
\label{general}

In this subsection, we study the four point blocks for \eqref{4pt0} but with the exchange of general operators. Using the expression in \eqref{Gnd} with \eqref{lw}, we evaluate the networks of open Wilson lines as
\begin{align}
\begin{aligned}
&G_{4,\text{e,o}}(j_i | z)  = \left. \left [ \prod_{i=1}^4 W_{j_i} (z_i ; z_0 ; X_i) \langle X_i | \right] | S \rangle_\text{e,o} \right|_{X_i = 0} \, .
\end{aligned}
\label{G41}
\end{align}
Here we set $j_1 = j_2 = q, j_3 = j_4 = j$ and $(z_1 , z_2 ,z_3 , z_4) = (z,0,\infty,1)$.
As explained in section \ref{Basics}, there are two independent superconformal blocks, where the integer and half-integer level of descendant operators are exchanged.

\begin{figure}
	\centering
	\includegraphics[keepaspectratio, scale=1]
	{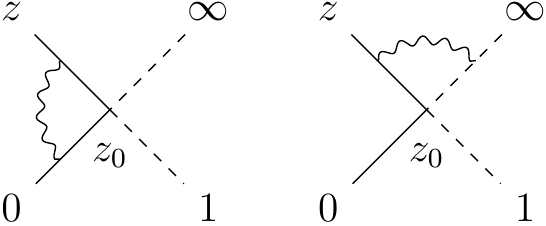}
	\caption{Contributions to the four point blocks from the networks of open Wilson lines at the $1/c$ order. The solid line represents an open Wilson line connecting $z$ and $0$ via $z_0$  and the dotted line represent an open Wilson line connecting $\infty$ and $1$ via $z_0$. 
		We should be careful when there is an exchange of spin $3/2$ field, and the situation depends on whether the exchange is between the solid and/or dotted lines.}
	\label{Wilson3}
\end{figure}

A main issue here is which solutions to the singlet equation \eqref{singlet} should be used.
With \eqref{leadingW}, the leading order expressions of superconformal blocks in $1/c$ are obtained  simply by replacing $x_i$ in the vertices with $z_i$. Therefore, we know which solutions are used at the leading order in $1/c$.
Moreover, as discussed at the end of subsection \ref{3ptblocks}, we may need to use a different vertex when the insertions of $G(z)$ are involved. 
In this subsection, we only consider the corrections of $1/c$ order, thus there should be either no insertion or two insertions of $G(z)$. The two insertions of $G(z)$ would change the type of $\langle VVVV \rangle $ into that with two $V$'s replaced by two $W$'s.
As explained in \eqref{ssSCB} in terms of superconformal invariants, there are two independent types which are not exchanged by superconformal transformations.
We can change the $z \to 0$ channel blocks for $\langle VVVV \rangle $ to those of 
\begin{align}
\langle WWVV \rangle \, , \quad \langle VVWW \rangle 
\label{WWVV}
\end{align}
by superconformal transformations.
However, this is not possible for 
\begin{align}
\langle WVWV \rangle \, , \quad \langle WVVW \rangle \, , \quad
\langle VWWV \rangle \, , \quad \langle VWVW \rangle \, .
\label{WVVW}
\end{align}
We call the vertices which reproduce the leading order expressions for $\langle VVVV \rangle $ (or, say, $\langle WVVW \rangle $) as those of type  $\langle VVVV \rangle $ (or $\langle WVVW \rangle $).
Our rule is to use the vertices of the type $\langle VVVV \rangle $, when there are no insertion of $G(z)$ or two insertions of $G(z)$ but in the open Wilson lines connecting $z_1$ and $z_2$ described by the solid lines or $z_3$ and $z_4$ described by the dotted lines in figure \ref{Wilson3}. 
Similarly, we  should use the vertices of the type  $\langle WVVW \rangle $, when there are two insertions of $G(z)$ in the open Wilson lines connecting $z_{1,2}$ and  $z_{3,4}$.

The leading order expressions of the four point blocks for \eqref{4pt0}  in $1/c$ were already given in \eqref{global0}. Thus the vertices of type $\langle VVVV \rangle $ can be obtained by replacing $z_i$ with $x_i$. For our application, it is convenient to rewrite them as the integration over the products of two three point blocks as (see \cite{Fateev:2011qa,Hikida:2018dxe})
\begin{align}
&v^{(1)}_{4, \text{e}} (j_i|X_i) = \mathcal{N}^{(1)}_\text{e} \int d x d \xi '  d\xi  \,  \xi \xi '  v_3^{(1)} (j_1 ,j_2 , p|X_1 ,X_2 ,X)
v_3^{(1)} (j_3 ,j_4 , -1-p|X_3 ,X_4 ,X ') \, ,  \nonumber \\
&v^{(1)}_{4, \text{o}}  (j_i|X_i)= \mathcal{N}^{(1)}_\text{o} \int d x d \xi '  d\xi   v_3^{(2)} (j_1 ,j_2 , p|X_1 ,X_2 ,X)
v_3^{(2)} (j_3 ,j_4 , -p|X_3 ,X_4 ,X ') 
\end{align}
with $x = x'$.
The integration contour is $x \in (0,\infty)$ and the prefactors are
\begin{align}
\mathcal{N}^{(1)}_\text{e} = \frac{(-1)^{p} \Gamma (-2 p)}{\Gamma (-p)^2} \, , \quad
\mathcal{N}^{(1)}_\text{o} = - \frac{1}{2p} \cdot  \frac{(-1)^{-\frac{1}{2}+p} \Gamma (1-2 p)}{\Gamma \left(\frac{1}{2}-p\right)^2} \, .
\end{align}
The leading order expressions of four point blocks for $\langle W V V W \rangle$ are computed in \eqref{global}. Similarly for the type  $\langle VVVV \rangle $, we use the expressions of vertices as
\begin{align}
&v^{(2)}_{4, \text{e}}  (j_i|X_i)= \mathcal{N}^{(2)}_\text{e} \int  d x d \xi '  d\xi  \,  \xi \xi '   v_3^{(2)} (j_1 ,j_2 , p|X_1 ,X_2 ,X)
v_3^{(2)} (j_3 ,j_4 , -1-p|X_3 ,X_4 ,X ') \, , \nonumber \\
&v^{(2)}_{4, \text{o}}  (j_i|X_i)= \mathcal{N}^{(2)}_\text{o} \int  d x d \xi '  d\xi   v_3^{(1)} (j_1 ,j_2 , p|X_1 ,X_2 ,X)
v_3^{(1)} (j_3 ,j_4 , -p|X_3,X_4,X ') 
\label{v42}
\end{align}
with $x = x'$.
The integration contour is $x \in (0,\infty)$ and the prefactors are%
\footnote{The first factor in $\mathcal{N}^{(1,2)}_o$ comes from the corresponding factors in \eqref{global0} and \eqref{global}. The second factor in  $\mathcal{N}^{(2)}_o$ arises from, say $\int d \xi ' d \xi \xi_1 \xi \xi_4 \xi '  = - \xi_1 \xi_4$.}
\begin{align}
\mathcal{N}^{(2)}_\text{e} = \frac{(-1)^{p+\frac{3}{2}} \Gamma (-2 p)}{\Gamma \left(-p-\frac{1}{2}\right) \Gamma \left(\frac{1}{2}-p\right)} \, , \quad
\mathcal{N}^{(2)}_\text{o} =  - \frac{p}{2}   \cdot (-1) \cdot \frac{ (-1)^{p} \Gamma (1-2 p)}{p \Gamma (1-p)^2} \, .
\label{v42n}
\end{align}

\begin{figure}
	\centering
	\includegraphics[keepaspectratio, scale=1]
	{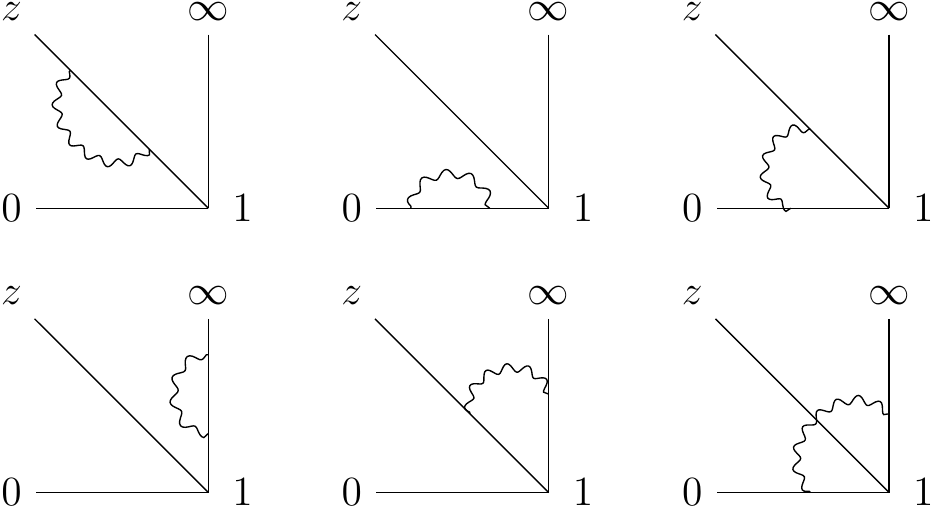}
	\caption{Contributions to the four point blocks from the networks of open Wilson lines at the $1/c$ order. The two bottom left diagrams are of the type as in the right diagram of figure \ref{Wilson3}  with $z_0 = 1$. Therefore we should be careful to choose the vertex when a spin $3/2$ current is exchanged.
	}
	\label{Wilson4}
\end{figure}

For the rest of computation, we closely follow the analysis in \cite{Hikida:2018dxe}.
The contributions to the $1/c$ corrections of four point blocks come from the diagrams in figure \ref{Wilson4}.
For the top diagrams, we compute
\begin{align}
&F^{(s)}_{\text{e},a} (z) = \mathcal{N}^{(1)} _\text{e} \int_0^\infty d x \, t^{s,\alpha=1}_a (z;x) x^{-1-p} \, , \\
&F^{(s)}_{\text{o},a} (z) = 
\mathcal{N}^{(1)} _\text{o} \int_0^\infty d x\,  t^{s,\alpha=2}_a (z;x) x^{-1/2-p} 
\end{align}
with $a=1,2,3$ and $s=2,3/2$, where $t^{s,\alpha}_a (z,x)$ were defined in \eqref{talpha} with $j_1 = j_2 = q$, $j_3 = p$.
We can show explicitly that the contributions from the bottom left diagram are proportional to the leading order expressions, so we neglect them.

For the other two diagrams, we define the functions
\begin{align}
\begin{aligned}
&u_s^{(\alpha)} (z ; z_a  ; y_2  ; x_3) \\
& \quad =  - \frac{6}{c} \frac{1}{N_s} \int d \xi_3 \eta_{s,\alpha} (\xi_3) \left.  V^s (z_a - y_1) e^{(z-1) V^2_1 (x_1)} e^{ - V^2_1 (x_2)}  v_3^{(\alpha)} (j_i | X_i) \right|_{X_{1,2}=0} 
\end{aligned}
\label{us}
\end{align}
with $V^s(z)$ in \eqref{Vs} and $j_1 = j_2 = q$, $j_3 = p$.
Here we have set
\begin{align}
\eta_{s,\alpha} (\xi_3) =
\begin{cases}
\xi_3 & (s, \alpha) = (2,1) \text{~ or ~} (3/2,2) \, ,\\
1 & (s, \alpha) = (2,2) \text{~ or ~} (3/2,1) \, .
\end{cases} 
\label{eta}
\end{align}
We further define the other functions
\begin{align}
\begin{aligned}
&\hat u_{s}^{(\alpha)} ( y_2  ; x_3) \\
& \quad =  - \frac{6}{c} \frac{1}{N_s} \lim_{z_\infty \to \infty} z_\infty^{-2 j} \int d \xi_3 \eta_{s,\alpha} (\xi_3) \left.  V^s (z_\infty - y_2) e^{(z_\infty-1) V^2_1 (x_1)}  v_3^{(\alpha)} (j_i | X_i) \right|_{X_{1,2}=0} 
\end{aligned}
\label{hatus}
\end{align}
with $j_1 = j_2 = j$ and  
\begin{align}
j_3 =
\begin{cases}
- 1 - p & (s, \alpha) = (2,1) \text{~ or ~} (3/2,2) \, ,\\
- p & (s, \alpha) = (2,2) \text{~ or ~} (3/2,1) \, .
\end{cases} 
\end{align}
Here $\eta_{s,\alpha} (\xi_3)$ is given in \eqref{eta}.
For the spin 2 exchanges, we evaluate the integrals
\begin{align}
\begin{aligned}
F_{\text{e},a+3}^{(2)} (z)  =&  \mathcal{N}^{(1)}_\text{e}   \int_0^\infty dx \int_1^\infty d y_2 \int_1^{z_a} d y_1  \\
& \qquad \qquad \times
u_2^{(1)} (z; z_a ; y_1 , x) \hat u_{2}^{(1)} (y_2 , x) \langle T(y_1) T(y_2) \rangle \, , \\
F_{\text{o},a+3}^{(2)} (z)  =& \mathcal{N}^{(1)}_\text{o}   \int_0^\infty dx \int_1^\infty d y_2 \int_1^{z_a} d y_1  \\
& \qquad \qquad \times u_2^{(2)} (z; z_a ; y_1 , x) \hat u_{2}^{(2)} (y_2 , x) \langle T(y_1) T(y_2) \rangle 
\end{aligned}
\end{align}
with $z_{a=1} =z$ and $z_{a=2}=0$. 
For the spin $3/2$ exchanges, we should use the vertices in \eqref{v42} with \eqref{v42n} as explained  above.
With the functions in \eqref{us} and \eqref{hatus},
the integrals to compute can be written as
\begin{align}
\begin{aligned}
F_{\text{e},a+3}^{(3/2)} (z)  =& 
\mathcal{N}^{(2)}_\text{o}  \int_0^\infty dx \int_1^\infty d y_2 \int_1^{z_a} d y_1  \\
& \qquad \qquad \times u_{3/2}^{(1)} (z; z_a ; y_1 , x) \hat u_{3/2}^{(1)} (y_2 , x) \langle G(y_1) G(y_2) \rangle \, , \\
F_{\text{o},a+3}^{(3/2)} (z)  =& 
\mathcal{N}^{(2)}_\text{e} \int_0^\infty dx \int_1^\infty d y_2 \int_1^{z_a} d y_1  \\
& \qquad \qquad \times u_{3/2}^{(2)} (z; z_a ; y_1 , x) \hat u_{3/2}^{(2)} (y_2 , x) \langle G(y_1) G(y_2) \rangle 
\end{aligned}
\end{align}
with $z_{a=1} =z$ and $z_{a=2}=0$.

The $1/c$ order contributions of the Wilson line networks in \eqref{G41} are given by the sums of all integrals as
\begin{align}
\left. \langle  G_\text{4,e,o} (j_i|z) \rangle \right|_{\mathcal{O}(c^{-1})}  = \sum_{a=1}^5 \sum_{s=2,3/2} F^{(s)}_{\text{e},a} (z) \, .
\end{align}
There could be terms proportional to the leading order expressions in the $1/c$ order contributions, but they can be subtracted (or tuned in a proper way). This is because we are interested in superconformal blocks, so the overall factors do not matter. Our normalizations are 
\begin{align}
z^{2 h_j - h_p} \langle  G_\text{4,e} (j_i|z) \rangle  = 1 + \sum_{n=1}^\infty a_{\text{e},n} z^n \, , \quad
z^{2 h_j - h_p - 1/2} \langle  G_\text{4,o} (j_i|z) \rangle  = \frac{1}{2h_p} + \sum_{n=1}^\infty  a_{\text{o},n} z^n
\end{align}
with 
\begin{align}
\frac{1}{2 h_p} = - \frac{1}{2p} + \frac{1}{c} \left( \frac{6 p + 3}{2 p  }\right) + \mathcal{O} (c^{-2}) \, .
\end{align}
In order to express the results, it might be useful to introduce functions as
\begin{align}
\begin{aligned}
z^{2 h_j - h_p} \langle  G_\text{4,e} (j_i|z) \rangle 
=  & {}_2 F_1 ( - p , - p ; - 2 p ;z) 
\\ &+ \frac{1}{c} \left[ q j f^\text{e}_a (p,z) + (q + j) f^\text{e}_b (p,z) + f^\text{e}_c ( p ,z) \right]  + \mathcal{O} (c^{-2}) \, , \\
z^{2 h_j - h_p - 1/2} \langle  G_\text{4,o} (j_i|z) \rangle 
=  & - \frac{1}{2p} {}_2 F_1 ( - p + \tfrac12 , - p + \tfrac12 ; - 2 p + 1 ;z) \\ & + \frac{1}{c} \left[ q j f^\text{o}_a (p,z) + (q + j) f^\text{o}_b (p,z) + f^\text{o}_c ( p ,z) \right] + \mathcal{O}(c^{-2})\, . 
\end{aligned}
\end{align}
We then find
\begin{align}
&f^\text{e}_a (p,z) = 12 ((z-2) \log (1-z)-2 z) \, _2F_1(-p,-p;-2 p;z) \, , \nonumber \\
&f^\text{e}_b (p,z) = 3 (4 p z \, _2F_1(-p,-p;-2 p;z)-4 p (z-1) \log (1-z) \, _2F_1(1-p,-p;-2 p;z)) \, , \nonumber \\
&f^\text{e}_c (p,z) = 3 (p \, _2F_1(-p,-p;-2 p;z) ((2 p+1) (z-1) \log (1-z)-2 p z)
\\ & \quad   -p (z-1) \log (1-z) \, _2F_1(1-p,-p;-2 p;z)) \, , \nonumber 
\end{align}
and
\begin{align}
&f^\text{o}_a (p,z) = -\frac{6}{p}  \left(((z-2) \log (1-z)-2 z) \, _2F_1\left(\tfrac{1}{2}-p,\tfrac{1}{2}-p;1-2 p;z\right)\right) \, , \nonumber \\
&f^\text{o}_b (p,z) = -\frac{3}{p} \left((2 p-1) \left(z \, _2F_1\left(\tfrac{1}{2}-p,\tfrac{1}{2}-p;1-2 p;z\right) \right. \right. \nonumber \\ & \quad \left. \left. -(z-1) \log (1-z) \, _2F_1\left(\tfrac{1}{2}-p,\tfrac{3}{2}-p;1-2 p;z\right)\right)\right) \, , \\
&f^\text{o}_c (p,z) =  \frac{3}{2 p z+z} \left(\, _2F_1\left(-p-\tfrac{1}{2},-p-\tfrac{1}{2};-2 p;z\right) \right. \nonumber \\ & \quad \left.  \times ((2 p+1) (2 p (z-1)-1) \log (1-z)    -(4 p (p+1)+3) z) \right.\nonumber  \\ & \quad \left.+\, _2F_1\left(-p-\tfrac{1}{2},\tfrac{1}{2}-p;-2 p;z\right) \left((4 p (p+1)+3) z-(2 p+1)^2 (z-1) \log (1-z)\right)\right)\, . \nonumber
\end{align}
We can check that they reproduce the $1/c$ order terms in \eqref{gendeg} by setting $q=1/2$. Moreover, the first few terms in $z$ expansions with general $h_i$ and finite $c$ were obtained in \cite{Belavin:2006zr} by solving recursion relations. 
We also find agreements for $h_1 = h_2 = h_q$, $h_3 = h_4 = h_j$, and $h = h_p$.

\section{Superconformal blocks involving heavy operators}

\label{heavy}

In the previous sections, we have considered light operators with conformal weight of order $c^0$.
In this section, we analyze the $1/c$ corrections for heavy operators with conformal weight of order $c$ from the bulk viewpoints.
We first construct supersymmetric conical spaces, which correspond to a set of heavy operators. 
We then examine the quantum corrections to the conical spaces by making use of the symmetry of the super Virasoro group, and we reproduce the $1/c$ corrections for the conformal weights of the corresponding heavy operators.
Using the analysis, we compute a heavy-heavy-light-light block from an open Wilson line in the conical geometry.

We begin with classical supersymmetric conical spaces in the osp$(1|2)$ Chern-Simons gauge theory.
In order to discuss the supersymmetric geometry, it is convenient to work with a cylindrical coordinate $w$, which is related to the planer one $z$ as $z = e^{i w}$.
With the coordinate, we use
\begin{align}
a (w) = V_{1}^2 - \frac{6}{c}  T(w) V_{-1}^2 - \frac{3 i^{1/2}}{c}  G (w) V^{3/2}_{-1/2}
\label{aw}
\end{align}
instead of \eqref{az}.
Asymptotic symmetry is given by the $\mathcal{N}=1$ superconformal algebra generated by
$T(w),G(w)$ \cite{Banados:1998pi} as mentioned in section \ref{Basics}. Here we put the phase factors with respect to $T(z)$ and $G(z)$ in the planar coordinate in order to match with the conventional notations.
We consider a gauge configuration independent of $w$ and corresponding to a geometry.
This implies that 
\begin{align}
T(w) = - \frac{ s^2 c }{2 4} \, .
\end{align}
Here any restriction is not yet assigned on $s$, and the condition for $s$ will be obtained by requiring the geometry preserving supersymmetry below. 

In order to obtain the supersymmetric condition, we apply the method developed for 
the supersymmetric geometry in the $\text{sl}(N+1|N)$ Chern-Simons theory as in \cite{Datta:2012km,Tan:2012xi,Hikida:2012eu,Chen:2013oxa,Datta:2013qja,Banados:2015tft}.
We look for spinors, which satisfy the Killing spinor equation, 
\begin{align}
{\cal D}_\mu \epsilon \equiv \partial_\mu \epsilon + [A_\mu , \epsilon] = 0 \, .
\end{align}
Formally, solutions can be written as
\begin{align}
\epsilon (x) = P \exp \left( - \int_{x_0}^x A_\mu dx^\mu \right) \hat \epsilon P \exp \left( \int_{x_0}^s A_\mu dx^\mu \right) 
\end{align}
with a constant spinor $\hat \epsilon $.
The problem is now whether the spinor satisfies the anti-periodic (or periodic) boundary condition in the NS-sector (or the R-sector). For the current configuration of gauge field, we find $(w = \phi + i \tau)$
\begin{align}
\exp \left( \oint d \phi A_\phi d \phi \right) = S^{-1} \exp \left( 2 \pi i  s V^2_{0} \right)S 
\end{align}
with a  matrix $S$. Rewriting the spinor as
\begin{align}
\epsilon  \equiv S \hat \epsilon  S^{-1} = \epsilon^- V^{3/2}_{-1/2} + \epsilon^+ V^{3/2}_{1/2} \, ,
\end{align}
the phase factor can be computed as
\begin{align}
\exp \left( - 2 \pi i s V^2_{0} \right) V^{3/2}_{\pm 1/2} \exp \left( 2  \pi i s V^2_{0} \right) 
= \exp \left( \pm \pi i s  \right) V^{3/2}_{\pm 1/2} \, .
\end{align}
Therefore, the anti-periodic (or periodic) boundary condition
leads to an odd (or even) integer $s$.
Recall that the conical defect geometry corresponds to the integer $s$ for the bosonic case \cite{Raeymaekers:2014kea}, see also \cite{Castro:2011iw}.

\subsection{Quantization of coadjoint orbits of super Virasoro group}

In order to analyze the quantum effects for the supersymmetric conical spaces, we utilize the  $\mathcal{N}=1$ super Virasoro symmetry generated by $T(w),G(w)$ in \eqref{aw}.
We use the description of the super Virasoro group in terms of superspace $(\tau , \theta)$ 
with $\tau \sim \tau + 2 \pi$
and super-derivative
\begin{align}
D_\theta = \partial_\theta + \theta \partial_\tau \, , \quad D_\theta ^2 = \partial_\tau \, .
\end{align}
Superconformal transformations are defined by \cite{Friedan:1986rx}
\begin{align}
\tau \to \tau ' (\tau , \theta) \, , \quad \theta \to \theta ' (\tau  ,\theta) 
\end{align}
subject to
\begin{align}
D \tau '  = \theta ' D \theta ' \, .
\end{align}
It will be convenient to use the parametrization (see, e.g., \cite{Fu:2016vas})
\begin{align}
\tau ' = \tau + f (\tau + \theta \eta (\tau)) \, , \quad
\theta ' = \sqrt{1 + \partial_\tau f(\tau)} \left[\theta + \eta (\tau) + \frac12 \theta \eta (\tau) \partial_\tau \eta (\tau) \right] \, ,
\end{align}
where $f(\tau)$ and $\eta (\tau)$ are bosonic and fermionic functions of $\tau$, respectively.
The computations will be done perturbatively in $f,\eta$ in the followings.

We examine the quantum effects of the geometry by applying quantization of the coadjoint orbits of the super Virasoro group. We follow the notation in \cite{Aoyama:1989pw}, see also \cite{Witten:1987ty,Alekseev:1988ce} for the case with the Virasoro group.
Elements of the super Virasoro algebra are given by pairs $(u(\tau, \theta),k)$, where $u(\tau,\theta)$ transforms as a weight $-1$ field and $k$ is a number. 
Dual elements are pairs $(\beta (\tau, \theta) , c)$, where $\beta(\tau,\theta)$ transforms as a weight $3/2$ field and $c$ is a number. The invariant quadratic form is defined as
\begin{align}
\langle (\beta ,c) , (u,k) \rangle = \frac{1}{2 \pi} \int_0^{2 \pi} d \tau d \theta \beta (\tau , \theta) u (\tau , \theta) + c k \, .
\end{align}
Considering an element of superconformal transformation $(\tau ,\theta) \to (\tau ' , \theta ')$, the coadjoint action Ad$^* (\tau ' , \theta ')$ can be expressed as
\begin{align}
\text{Ad}^* (\tau ' , \theta ') (\beta (\tau , \theta) ,c) 
= \left( \beta (\tau ' , \theta ') (D \theta ')^3 - \frac{c}{6} S(\theta ') , c \right)
\end{align}
with the super-Schwarzian derivative \cite{Friedan:1986rx}
\begin{align}
S(\theta ') = \frac{D^4 \theta '}{ D \theta '} - 2 \frac{D^3 \theta '}{D \theta '} \frac{D^2 \theta '}{ D \theta '} \, .
\end{align} 
A group element on 
the coadjoint orbit including a point $(\beta_0 (x , \theta) , x)$ can be expressed as
\begin{align}
(\beta (\tau, \theta) , c)^{\tau ' , \theta '} = \text{Ad}^* (\tau ' , \theta ') (\beta_0 (\tau , \theta) ,c) \, .
\end{align}
With this parametrization, the Kirillov-Kostant symplectic form was found to be \cite{Aoyama:1989pw}
\begin{align}
\begin{aligned}
\Omega
= \frac{1}{2 \pi}d \int_0^{2 \pi} d \tau d \theta \Biggl[ & \beta_0 (\tau ' , \theta ') (d \tau ' + \theta ' d \theta ') D \theta '  \\ &
- \frac{c}{12} \frac{d \tau ' + \theta ' d \theta '}{(D \theta ')^2} 
\left( \frac{D^4 \theta '}{D \theta '} - 3 \frac{D^3 \theta '}{D \theta '} \frac{D^2 \theta '}{D \theta '}\right)  \Biggr]\, .
\end{aligned}
\label{KK2form}
\end{align}
We quantize the coadjoint orbits by using the symplectic form.

We consider the coadjoint orbit including the point
\begin{align}
\beta (\tau , \theta) = - \frac{c s^2}{24} \theta \, .
\end{align}
Notice that $\beta (\tau , \theta)$ corresponds to a superfield $T(\tau , \theta) = T_F (\tau) + \theta T_B (\tau)$ with $G(\tau) = 2 T_F (\tau) $, $T(\tau) = T_B(\tau) $ introduced in section \ref{Basics}.
We choose a basis for super Virasoro generators as
\begin{align}
&\ell_n =\langle ( \theta e^{i n \tau} , 0) , (\beta (\tau, \theta) , c)^{\tau ' , \theta '}\rangle = \frac{1}{2 \pi} \int_0^{2 \pi} d \tau d \theta \theta e^{i n \tau} 
\left( - \frac{c s^2 }{24} \theta '  (D \theta ')^3- \frac{c}{6} S ( \theta ') \right) \, ,\\
&g_r = 2 \langle ( e^{i r \tau} , 0) , (\beta (\tau, \theta) , c)^{\tau ' , \theta '}\rangle = \frac{1}{ \pi} \int_0^{2 \pi} d \tau d \theta  e^{i r \tau} \left( - \frac{c s^2 }{24} \theta ' (D \theta ')^3 - \frac{c}{6} S ( \theta ') \right) \, ,
\end{align}
where $r \in \mathbb{Z} + 1/2$ for the NS-sector $(s \in 2 \mathbb{Z} +1)$ and $r \in \mathbb{Z}$ for the R-sector $(s \in 2 \mathbb{Z})$.
In terms of mode expansions as
\begin{align}
f(\tau) = \sum_n f_n e^{- i n \tau} \, , \quad
\eta (\tau) = \sum_{r} \eta_r e^{- i r \tau} \, ,
\end{align}
we find
\begin{align}
&\ell_0 = - \frac{c s^2}{24} + \frac{c}{24} \sum_n n^2 (n^2 - s^2) f_n f_{-n}
-  \frac{ic}{6} \sum_{r} r \left(r^2 - \left(\frac{s}{2} \right)^2 \right) \eta_{r} \eta_{-r} \, , \\
&\ell_{\pm s} = \frac{c}{24} \sum_n (n^2 - s^2 ) \left((n-s) ^2 - s^2 \right) f_n f_{- n \pm s } 
- \frac{i c}{6} \sum_r \left(r \mp \frac{3}{2}s \right) \left(r \mp \frac{s}{2} \right) \left(r \pm \frac{s}{2}\right) \eta_r \eta_{-r \pm s} \, , \nonumber \\
& g_{\pm s/2} =  - \frac{ic}{6} \sum_{n} n (n^2 - s^2) f_n \eta_{- n \pm s/2} \nonumber
\end{align}
up to the terms including two of $f_n, \eta_r$.
Similarly, we obtain
\begin{align}
\ell_m = - \frac{i c}{12} m (m^2 - s^2 ) f_m \, , \quad
g_r =  \frac{c}{3} \left(r^2 - \left(\frac{s}{2} \right)^2  \right) \eta_r 
\end{align}
for $m \neq 0 , \pm s$ and $r \neq \pm s/2$ up to the terms including one of $f_n, \eta_r$.
Note that the above expressions are independent of $f_0 ,f_{\pm s} , \eta_{\pm s/2}$.

The Kirillov-Kostant symplectic form in \eqref{KK2form} can be computed as
\begin{align}
\Omega = \frac{c}{12} \left[ - i \sum_{m >0} m( m^2 -  s^2 ) df_m d f_{-m} -  \sum_{r >0} ( 4 r^2 - s^2 ) d \eta_r d \eta_{-r} \right] 
\label{KK2form2}
\end{align}
up to the terms with two of $f_n, \eta_r$.
Let us first consider the NS-sector with $r \in \mathbb{Z} + 1/2$.
The Poisson (anti-) brackets are, then, given by
\begin{align}
- i \{ f_m , f_n  \}_\text{PB} = \frac{12 }{c} \frac{ \delta_{m+n}}{m(m^2 - s^2)} \, , \quad
- i \{ \eta_r , \eta_u \}_\text{PB} = \frac{3}{c} \frac{ i \delta_{r+u}}{r^2 - (s/2)^2} 
\label{PB}
\end{align}
up to the current order.
Therefore, the quantization is possible by replacing
\begin{align}
\sqrt{\frac{c}{12}} f_m \to A_m \, , \quad \sqrt{\frac{c}{3}} \eta_r \to B_r
\end{align}
with (anti-)commutation relations 
\begin{align}
[A_m , A_n] = \frac{\delta_{m+n}}{m(m^2 -s^2)} \, , \quad
\{B_r , B_u \} = \frac{i \delta_{r + u}}{r^2 - (s/2)^2} \, .
\label{AB}
\end{align}
This implies that the expansion in terms of $f_r ,\eta_r$ corresponds to the quantization with the Planck constant as $h \sim 1/c$, see \cite{Witten:1987ty} for the Virasoro group.
For the R-sector, we use
\begin{align}
- i \{ \eta_0 , \eta_0 \}_\text{PB} = \frac{1}{2} \frac{3}{c} \frac{- i }{  (s/2)^2} \, , \quad
\{ B_0 , B_0 \} = \frac{1}{2}  \frac{- i }{  (s/2)^2} 
\label{PB2}
\end{align}
in addition to \eqref{PB} and \eqref{AB} expect for those involving $\eta_0$ and $B_0$.
The vacuum state and its dual are defined as
\begin{align}
A_m | 0 \rangle_s = 0 \, , \quad B_r | 0 \rangle_s = 0 \, , \quad
{}_s \langle  0 | A_{-m}= 0 \, , \quad {}_s \langle 0 |  B_{-r} = 0 
\end{align}
for $m > 0$ and $r > 0$.

When we replace $f_m ,\eta_r$ by operators $A_m ,B_r$, we should take care of the ambiguity related to the ordering of operators. This ambiguity exists for $\ell_0 \to \hat \ell_0$, and we express it as 
\begin{align}
&\hat \ell_0 = - \frac{c s^2}{24} + n_0 +  \sum_{n > 0} n^2 (n^2 - s^2 ) A_n A_{-n}
- i  \sum_{r > 0} r (r^2 - (s/2)^2 ) B_{r} B_{-r} 
\end{align}
by introducing a parameter $n_0$. Here $n_0$ corresponds to the $1/c$ correction for the conformal weight of the heavy operator dual to the supersymmetric conical geometry with label $s$.
As in \cite{Raeymaekers:2014kea}, we fix $n_0$ by requiring that the generators satisfy the (anti-)commutation relations of the superconformal algebra. 
Using 
\begin{align}
\hat g_{\pm s/2} = - i \sum_n n (n^2 - s^2 ) A_n B_{- n \pm s/2} \, , 
\end{align}
we  require that%
\footnote{We should modify the anti-commutation relation in \eqref{commutators} due to the map from the planer coordinate to the cylindrical one.  The phase factors in \eqref{Fs} arises just like for $T(w), G(w)$ in \eqref{aw}. Moreover the shift $L_0 \to L_0 - \frac{c}{24}$ removes the term $- \frac{c}{24} \delta_{r+t}$ in \eqref{commutators}.}
\begin{align}
{}_s \langle 0 | \mathcal{F}_{s} | 0 \rangle_s = 0 \, ,  \quad 
\mathcal{F}_{s} \equiv- i \{\hat g_{s/2} , \hat g_{-s/2} \} - 2 \hat \ell_{0} - \frac{c}{3} \frac{s^2}{4} \, . \label{Fs}
\end{align}
This, in particular, means 
\begin{align}
n_0 =  - \frac{  i }{2} {}_s \langle 0 |\{\hat g_{s/2} , \hat g_{-s/2} \}  | 0 \rangle_s \, .
\label{n0}
\end{align}
Using the  (anti-)commutation relation, we find
\begin{align}
{}_s \langle 0 |\{\hat g_{s/2} , \hat g_{-s/2} \}  | 0 \rangle_s = i
\sum_{n=1}^{(s-1)/2} \frac{n (n^2 - s^2)}{(n - s/2)^2 - (s/2)^2} = \frac{i}{8} (5s^2 - 4 s - 1) 
\end{align}
for odd $s$ and
\begin{align}
{}_s \langle 0 |\{\hat g_{s/2} , \hat g_{-s/2} \}  | 0 \rangle_s =  i 
\sum_{n=1}^{s/2- 1 } \frac{n (n^2 - s^2)}{(n - s/2)^2 - (s/2)^2} 
+\frac{i}{2} \frac{3 s}{2}  = \frac{i}{8} s (5s - 4 ) 
\end{align}
for even $s$. From \eqref{n0}, we arrive at the result
\begin{align}
n_0 = \frac{1}{16} (5s^2 - 4 s - 1) + \frac{1 + (-1)^s}{32} \, , 
\end{align}
which correctly reproduces the $c^0$ order term in \eqref{h1s}.

\subsection{Open Wilson line in a supersymmetric conical space}

In our previous work \cite{Hikida:2018dxe}, we have developed a way to compute a heavy-heavy-light-light block from the bulk perspective, and
in this subsection, we generalize the analysis to the supersymmetric case.
As a heavy-heavy-light-light block, we consider
\begin{align}
\mathcal{G}_s (z) = \langle V_{h_{3,1}} (z) V_{h_{3,1}} (0) V_{h_{1,s}} (\infty) V_{h_{1,s}} (1)\rangle \, ,
\label{gs}
\end{align}
whose expression in $z$ expansion is obtained in appendix \ref{CFT}.
In the cylindrical coordinate, the block can be expressed as
\begin{align}
\tilde{\mathcal{G}}_s (w) = \langle V_{h_{3,1}} (w) V_{h_{3,1}} (0) V_{h_{1,s}} (- i \infty) V_{h_{1,s}} (i \infty)\rangle \, ,
\end{align}
which is related to \eqref{gs} as
\begin{align}
\mathcal{G}_s (z) =- i ( 1-z)^{- h_{3,1}} \left.  \tilde{\mathcal{G}}_s (w) \right|_{e^{i w} = 1 -z} \, .
\label{gszgsw}
\end{align}
The phase factor is included for later convenience.

Following the analysis in \cite{Hikida:2018dxe}, we compute the superconformal block from an open Wilson line in a supersymmetric conical geometry with label $s$ as
\begin{align}
W_s (w) = P \exp \left[ \int_0^w d w ' \left( V_{1}^2 - \frac{6}{c} T(w ') V_{-1}^2 - \frac{3 i^{1/2}}{c}  G (w ') V^{3/2}_{-1/2} \right)\right] \, ,
\end{align}
where we have used \eqref{aw}.
We examine the Wilson line operator in $1/c$ expansion.
We define
\begin{align}
\hat J^{(2)}( w ) = T(w) + \frac{s^2 c}{24} \, , \quad
\hat J^{(3/2)} (w) = G (w)  
\end{align}
such that the two point functions behaves of order $c$ (or $\hat J^{(s)} (w)$ behaves of order $c^{1/2}$) as in (5.27) of \cite{Hikida:2018dxe} and \eqref{2ptcurrentsNS}, \eqref{2ptcurrentsR} below. 
We rewrite the Wilson line operator as
\begin{align}
\begin{aligned}
W_s (w) 
= & \sum_{n=0}^\infty \frac{1}{c^n} \int_0^w d w_n \cdots \int_0^{w_2} d w_1 \\
& \qquad \times \sum_{t_j = 3/2,2}
\left[ \prod_{j=1}^n \mathcal{C}_{t_j} \hat J^{(t_j)} (w_j) \right] f_s^{(t_n , \ldots , t_1)} (w_n , \ldots , w_1) 
\end{aligned}
\label{Ws}
\end{align}
with 
\begin{align}
\begin{aligned}
&f_s^{(t_n , \ldots , t_1)} (w_n , \ldots , w_1) 
= (V^{t_n})_s (w_{*n}) \cdots (V^{t_1})_s (w_{*1}) \left. e^{(V_1^2 + \frac{s^2}{4} V_{-1}^2)w }x \right|_{x=0} \, , \\
&(V^t)_s (w) \equiv e^{(V_1^2 + \frac{s^2}{4} V_{-1}^2) w} V^{t}_{-t+1} e^{-(V_1^2 + \frac{s^2}{4} V_{-1}^2) w} \, .
\end{aligned}
\end{align}
Here we have used
\begin{align}
\mathcal{C}_2 = - 6 \, , \quad \mathcal{C}_{3/2} = - 3  i^{1/2} \, , \quad w_{*i} = w - w_i \, .
\end{align}
We set $c_s=1$ in \eqref{Ws} since we consider only the next leading contributions in $1/c$.
The leading order in $1/c$ becomes
\begin{align}
W_s (w) = \frac{2}{s} \sin \left( \frac{s w}{2} \right) + \mathcal{O} (c^{-1})\, ,
\end{align}
which reproduces the $z$ expansion of the leading order term of \eqref{gsleading} in $1/c$ by using \eqref{gszgsw}. In fact, it is the same as the bosonic case in \eqref{Gsexp},
see \cite{Chen:2016cms} for general arguments.

We then move to the $1/c$ corrections.
The contributions can be divided into two types; one is with the insertions of one current and the other with two currents.
A type of $1/c$ correction with a current insertion comes from
\begin{align}
\begin{aligned}
n_0 M_s^{(1)} (w)&= - \frac{6}{c} \int_0^w d w_1 f_s^{(2)} (w_1) {}_s \langle 0 | \hat J^{(2)} (w_1) | 0 \rangle_s  \\
&= \frac{12 n_0}{c} \left[ \frac{2 \sin \left( \frac{s w}{2}\right) - s w \cos \left(\frac{s w}{2} \right)}{s^3} \right] \, .
\end{aligned}
\end{align}
This is almost the same as that in the bosonic case but with a different value of $n_0$.
The difference is
\begin{align}
\begin{aligned}
\Delta n_0 &= \frac{(5 s + 1) (s - 1)}{16 } + \frac{1 + (-1)^s}{32} - \frac{(s - 1) (1 + 13 s) }{24}
\\ &= -\frac{(s-1) (11 s-1)}{48} + \frac{1 + (-1)^s}{32} \, .
\end{aligned}
\end{align}
Another contribution is with the insertion of two $\hat J^{(2)}(w)$'s, but it is exactly the same as the bosonic case as
\begin{align}
M_s^{(2)} (w)= \left( - \frac{6}{c}\right)^2   \int_0^w d w_2 \int_0^{w_2} d w_1 f_s^{(2,2)} (w_2 ,w_1) {}_s \langle 0 | \hat J^{(2)} (w_2) \hat J^{(2)} (w_1) | 0 \rangle_s \, .
\end{align}
Therefore, we do not repeat here.
There is also a contribution with the insertion of two $\hat J^{(3/2)}(w)$'s (or two $G(w)$'s), which is new in the current case.
Below we examine the integral for the contribution.

We thus evaluate
\begin{align}
M_s^{(3)} (w)= \left( - \frac{3 i^{1/2}}{c}\right)^2 \int_0^w d w_2 \int_0^{w_2} d w_1 f_s^{(3/2,3/2)} (w_2 ,w_1) {}_s \langle 0 | G(w_2) G (w_1) | 0 \rangle_s 
\label{Is3}
\end{align}
with
\begin{align}
f_s^{(3/2,3/2)} (w_2 ,w_1) = - \frac{4}{s^2} \sin \left(\frac{s w_1}{2}\right) \sin \left(\frac{s (w-w_2)}{2} \right) \, .
\end{align}
Let us first set $s$ to be odd and then discuss the case with even $s$.
We define the two point function of the superconformal current as
\begin{align}
E_s(w_2 ,w_1) \equiv
{}_s \langle 0 | G(w_2) G (w_1) | 0 \rangle_s \, ,
\label{2ptcurrents}
\end{align}
which is computed as
\begin{align}
\begin{aligned}
E_s(w_2 ,w_1)
= \sum_{r > 0} e^{- i r w_{21}} \frac{c (4 r^2 - s^2)}{12}
= \frac{i c}{12} \left( \frac{1}{\sin ^3 \left( \frac{w_{21}}{2} \right)} + \frac{s^2 -1}{2 \sin \left( \frac{w_{21}}{2} \right) }\right)
\label{2ptcurrentsNS}
\end{aligned}
\end{align}
with $w_{21} = w_2 - w_1$.
The integral \eqref{Is3} diverges due to the singular behavior of the two point function in \eqref{2ptcurrentsNS}.
As in the bosonic case analyzed in \cite{Fitzpatrick:2015dlt,Hikida:2018dxe}, 
we redefine the integral \eqref{Is3} by removing the divergent term as
\begin{align}
\begin{aligned}
M_{s,\text{reg}}^{(3)} (w)=& \frac{9 i x^{-s/2}}{ s^2 c}\int_0^x dx_2 \int_0^{x_2} dx_1
x_1^{s-1} x_2^{s-1} \left(1- x_1^s\right)  \left(x_2^s-x^s\right) \\
&\times \left(\frac{2 (x_1 x_2)^{\frac{3}{2}-\frac{3 s}{2}}}{3 (x_2- x_1)^3} -\frac{\left(s^2-1\right) (x_1 x_2)^{\frac{1}{2}-\frac{3 s}{2}}}{12 (x_2-x_1)} -\frac{2 s^3}{3 \left(x_2^s- x_1^s\right)^3}\right) \, .
\end{aligned}
\end{align}
Here we have introduced new parameters as
\begin{align}
x =e^{i w} \, , \quad x_1 = e^{i w_1} \, , \quad x_2 = e^{i w_2} \, , \quad
y = e^{i s w} \, , \quad y_1 = e^{i s w_1} \, , \quad y_2 = e^{i s w_2} \, .
\end{align}
The first few examples for the integral are
\begin{align}
& \frac{s^2 c} {9 i x^{-s/2}}M^{(3)}_{3 , \text{reg}} (w) = \frac{1}{27} \left(\left(51 x^3-3\right) \log (x)-27 \left(x^3-1\right) \log \left(x^2+x+1\right) \right. \nonumber \\ & \qquad  \left. +x (x (-25 x+27 x \log (3)+27)-27)+25-27 \log (3)\right) \, , \nonumber   \\
&\frac{s^2 c} {9 i x^{-s/2}}M^{(3)}_{5 , \text{reg}}(w) =\frac{1}{3} \left(-\frac{71 x^5}{10}+5 x^5 \log (5)+\left(19 x^5-1\right) \log (z)+5 x^4+\frac{5 x^3}{2} \right. \nonumber  \\ & \qquad \left. -\frac{5 x^2}{2} -5 \left(x^5-1\right) \log \left(x^4+x^3+x^2+x+1\right)-5 x+\frac{71}{10}-5 \log (5)\right) \, ,  \\
&\frac{s^2 c} {9 i x^{-s/2}} M^{(3)}_{7, \text{reg}}  (w)= \frac{1}{126} \left(-515 x^7+294 x^7 \log (7) \right.\nonumber    \\ & \qquad+84 \left(20 x^7-1\right) \log (x)  +294 x^6+147 x^5  +98 x^4-98 x^3  -147 x^2   \nonumber  \\ & \qquad \left. -294 \left(x^7-1\right) \log \left(x^6+x^5+x^4+x^3+x^2+z+1\right)-294 x+515-294 \log (7)\right) \, . \nonumber 
\end{align}
The divergent integral is
\begin{align}
M_{s,\text{div}}^{(3)} (w) = \frac{6  i y^{-s/2 }}{ s c} \int_1^y d y_2 \int_1^{y_2} dy_1 \frac{(1-y) (y_2-y)}{(y_2-y_1)^{3-2 \epsilon}} \, .
\end{align}
Taking care of the coordinate transformation from $y$ to $w$ as in \cite{Hikida:2018dxe}, we  use
\begin{align}
M_{s,\text{div}}^{(3)} (w) =\frac{3  i  e^{-i s w /2} (e^{is w} -1) }{ s c} \left[ \log (e^{i s w} - 1)   - \frac{1}{2} \log (e^{is w}) + \delta \right] \, .
\label{div}
\end{align}
Here a new parameter $\delta$ is introduced by using the ambiguity in the overall normalization at the order of $\epsilon^0$.

Now we compare the result obtained above to  \eqref{gsnext} computed with the technique of superconformal field theory.
As explained above, some of the contributions are the same as those for the bosonic case and the others are not. The additional part is given by the sum as
\begin{align}
\Delta n_0 M_s^{(1)} (w) + M_{s,\text{reg}}^{(3)} (w)+ M_{s,\text{div}}^{(3)} (w)\, .
\end{align}
Noticing \eqref{gszgsw},
we can confirm that this reproduces the expression of \eqref{gsnext} minus the $1/c$ order term in \eqref{Gs} or in \eqref{Gsexp} with  $\delta = - \log(s)$.%
\footnote{We have checked this for $s=3,5,\ldots,41$.}

We can similarly analyze the even $s$ case. For this case, the two point function of the superconformal current in \eqref{2ptcurrents} is computed as
\begin{align}
\begin{aligned}
E_s(w_2 ,w_1) 
= \sum_{r =1}^\infty e^{- i r w_{21}} \frac{c (4 r^2 - s^2)}{12}  -\frac{c s^2}{24}
= \frac{i c }{12}  \left( \frac{\cos \left(\frac{w_{21}}{2}\right) }{\sin ^3  \left(\frac{w_{21}}{2}\right) } +  \frac{ s^2  \cos \left(\frac{w_{21}}{2}\right) }{2 \sin \left(\frac{w_{21}}{2}\right) } \right) \, .
\end{aligned}
\label{2ptcurrentsR}
\end{align}
Just as in the odd $s$ case, we define the integral by removing the divergent term as
\begin{align}
\begin{aligned}
M_{s,\text{reg}}^{(3)} (w)=& \frac{9 i x^{- s/2}}{ s^2 c}\int_0^x dx_2 \int_0^{x_2} dx_1
x_1^{s-1} x_2^{s-1} \left(1- x_1^s\right)  \left(x_2^s-x^s\right) \\
&\times \left(\frac{ (x_1 x_2)^{1 - \frac{3s}{2}} \left(x_1 + x_2\right) }{3 \left(x_2 - x_1 \right)^3} - \frac{ s^2 (x_1 x_2)^{- \frac{3s}{2}} \left(x_1 + x_2\right) }{24 \left(x_2 - x_1 \right)}  -\frac{2 s^3}{3 \left(x_2^s- x_1^s\right)^3}\right) \, .
\end{aligned}
\end{align}
We can perform the integrals for specific $s=2,4,6,\cdots$ with no difficulty.
The evaluation of the divergent integral is the same as in \eqref{div}.
In this way, we can compute the $1/c$ correction of the block with specific $s=2,4,6,\cdots$,
though we do not go into the details here.

\section{Conclusion and open problems}
\label{conclusion}

In this paper, we computed the $\mathcal{N}=1$ superconformal blocks form the networks of open Wilson lines in the osp$(1|2)$ Chern-Simons theory in the large $c$ expansion. This is a supersymmetric extension of the previous works in \cite{Hikida:2017ehf,Hikida:2018dxe}, and one of the aim is to show that the method developed in these papers is useful for examining other models as well.
There were, however, several points we should elaborate in the current example as explained below.

We reproduced the conformal weight of light operator in \eqref{hexp} from an open Wilson line up to the $1/c^2$ order. For the purpose, we fixed the parameter $c_s$ introduced in \eqref{Wilsonopr}
by making use of the superconformal Ward-Takahashi identities \eqref{WT}. Since the insertion of $G(z)$ exchanges $V_h(z)$ and $\Lambda_h(z)$,  we should take care of the renormalization of the two operators $Z_V,Z_W$ in terms of open Wilson line as in \eqref{3ptf}.
We also computed the $1/c$ corrections of superconformal blocks from the networks of open Wilson lines. At the connecting point $z_0$, we should assign the vertex $| S \rangle$, which is given by a solution to the singlet conditions \eqref{singlet}. There are several solutions, and we may need to use several ones, even when we consider only one type of superconformal blocks. This is because there could be the insertions of $G(z)$ from open Wilson lines, and they can change the vertex we should use, see figure \ref{Wilson3}.
Furthermore, we examined the $1/c$ corrections associated with a heavy operator from a supersymmetric conical space. A main point is to construct the symplectic form over the coadjoint orbits of the super Virasoro group in order to quantize the geometry. Fortunately, it was already obtained in \cite{Aoyama:1989pw}. Applying the result, the rest is rather straightforward by extending the analysis in \cite{Raeymaekers:2014kea} for the conformal weight and our previous one in \cite{Hikida:2018dxe} for a heavy-heavy-light-light block.

We would like to think about the following open problems.
For light operators, we considered only the NS-sector, and it would be interesting to deal with also the R-sector. In particular, the correlators with the simplest degenerate operator in the R-sector satisfy the second order differential equations \cite{Fukuda:2002bv}. Therefore, there would be a simple way to express light operators in the R-sector even in terms of the osp$(1|2)$ Chern-Simons gauge theory. For heavy operators, we analyzed the R-sector as well in terms of Wilson line operator in conical geometry. We have not compared the results with those from conformal field theory, but it should not be so difficult to do so. We would like to systematically analyze correlators involving  operators in the R-sector from the viewpoint of superconformal field theory. 
In our previous paper \cite{Hikida:2018dxe}, we studied only a simple example of general W$_3$ four point block from the bulk viewpoints, and the current work may be useful to extend the analysis for more general blocks.

In this paper, we considered the $\mathcal{N}=1$ super Virasoro minimal model since it is the simplest example with supersymmetry. However, the examples with extended supersymmetry are physically more interesting as mentioned in the introduction.
For examples, we could similarly analyze the $\mathcal{N}=p$ superconformal blocks with $p=2,3,4$ in terms of Chern-Simons gauge theory as in \cite{Henneaux:2010xg}.
We also would like to study the blocks with respect to the $\mathcal{N}=2$ super W$_{N+1}$ algebra and their extensions.
These superconformal blocks should be useful  as fundamental objects for the study of superstring theory and AdS/CFT correspondence.
Furthermore, we could read off higher spin charges in the $\mathcal{N}=2$ W$_{N+1}$ minimal model by using the $\mathcal{N}=2$ superconformal blocks as in \cite{Hikida:2017byl}. 
It might be rather easier to use the $\mathcal{N}=1$ superconformal blocks obtained here in the $\mathcal{N}=1$ higher spin holographies of \cite{Creutzig:2012ar,Beccaria:2013wqa}.
Concerning to the $1/c$ corrections for heavy operators, we have utilized the quantization of the coadjoint orbits of the super Virasoro group. It is worthwhile to generalize the analysis for the cases with extended supersymmetry and/or higher spin symmetry, see \cite{Campoleoni:2017xyl} for a different approach with free-field variables.

\subsection*{Acknowledgements}

We are grateful to Thomas Creutzig, Yuya Kusuki, Henry Maxfield, Yuji Sugawara, Tadashi Takayanagi, and Satoshi Yamaguchi  for useful discussions. 
The work of YH is supported by JSPS KAKENHI Grant Number 16H02182.

\appendix

\section{$\mathcal{N}=1$ superconformal algebra and correlators}
\label{N1WT}

In this appendix, we summarize some useful results on the $\mathcal{N}=1$ superconformal algebra.
The superconformal generators satisfy the operator product expansions 
\begin{align}
\begin{aligned}
&T(z) T(w) \sim \frac{c/2}{(z-w)^4} + \frac{2 T(w)}{(z-w)^2} + \frac{\partial T(w)}{z-w} \, , \\
&T(z) G(w) \sim \frac{3/2 G(w)}{(z-w)^2} + \frac{\partial G(w)}{z-w} \, , \quad
G(z) G(w) \sim \frac{2 c /3 }{(z - w)^3} + \frac{2 T(w)}{z-w} \, .
\end{aligned}
\end{align}
The correlators of the superconformal currents follow the above relations, and in the main context, we have used those with the shifts of conformal weights as \eqref{reg} and
\begin{align}
\begin{aligned}
&\langle T(z_3) T(z_2) T(z_1) \rangle = \frac{c}{z_{32}^{2-\epsilon}z_{21}^{2-\epsilon}z_{31}^{2-\epsilon}} \, , \quad
\langle T(z_3) G(z_2) G(z_1) \rangle = \frac{c}{z_{32}^{2-\epsilon}z_{21}^{1-\epsilon}z_{31}^{2-\epsilon}} \, , \\
&\langle T(z_4) T(z_3) T(z_2) T(z_1) \rangle = \frac{(c/2)^2}{z_{43}^{4-2\epsilon}z_{21}^{4-2\epsilon}} + \frac{(c/2)^2}{z_{42}^{4-2\epsilon}z_{31}^{4-2\epsilon}} + \frac{(c/2)^2}{z_{41}^{4-2\epsilon}z_{32}^{4-2\epsilon}}  + \mathcal{O}(c)\, , \\
&\langle T(z_4) T(z_3) G(z_2) G(z_1) \rangle = \frac{c/2\cdot2c/3}{z_{43}^{4-2\epsilon}z_{21}^{3-2\epsilon}} + \mathcal{O}(c) \, , \\
&\langle G(z_4) G(z_3) G(z_2) G(z_1) \rangle = \frac{(2c/3)^2}{z_{43}^{3-2\epsilon}z_{21}^{3-2\epsilon}} - \frac{(2c/3)^2}{z_{42}^{3-2\epsilon}z_{31}^{3-2\epsilon}} + \frac{(2c/3)^2}{z_{41}^{3-2\epsilon}z_{32}^{3-2\epsilon}}+ \mathcal{O}(c) 
\end{aligned}
\label{regulator}
\end{align}
at the leading order in $1/c$.

We have also made use of the superconformal Ward-Takahashi identities in \eqref{WT}, and we give a derivation of the right equation here.
Using $\Phi_h (z , \theta) = V_h (z) + \theta \Lambda_h (z)$,
we assume
\begin{align}
\begin{aligned}
&T(z) V_h (0) \sim \frac{h V_h (0)}{z^2} + \frac{\partial V_h (0)}{z} \, , \quad
T(z ) \Lambda_h (0) \sim \frac{(h+\frac12) \Lambda_h (0)}{z^2} + \frac{\partial \Lambda_h (0)}{z} \, , \\
&G(z) V_h (0) \sim \frac{\Lambda_h (0)}{z} \, , \quad
G(z) \Lambda_h (0) \sim \frac{2 h V_h (0)}{z^2 } + \frac{\partial V_h (0)}{z} \, . 
\end{aligned}
\end{align}
The superconformal Ward-Takahashi identity can be derived as
\begin{align}
\begin{aligned}
\langle G(z) \Lambda_h (y) V_h (x) \rangle 
&= \left( \frac{2 h}{(z - y)^2} + \frac{1}{z -y}\partial_y \right) \langle V_h (y) V_h (x) \rangle
- \frac{1}{z - x} \langle \Lambda_h (y) \Lambda_h (x) \rangle 
\\ &= \frac{2h} {(x -z ) ( y -z)^2 (x - y)^{2 h -1}} \, .
\end{aligned}
\end{align}
Here we have used
\begin{align}
\langle V_h (y) V_h (x) \rangle = \frac{1}{(y - x)^{2h}} \, , \quad
\langle \Lambda_h (y) \Lambda_h (x) \rangle = - \frac{ 2 h}{(y - x)^{2h}} \, .
\end{align}

\section{Four point blocks in superconformal field theory}
\label{4ptblocks}

In this appendix, we examine superconformal blocks using the standard techniques of superconformal field theory.

\subsection{Superconformal blocks at the large $c$ limit}
\label{GCB}

In the main context, we have computed the four point blocks of the type $\langle VVVV \rangle$ including $1/c$ corrections from the networks of open Wilson line. At the connecting point, we need to use a vertex satisfying the singlet condition in \eqref{singlet}. For the leading order in $1/c$, we choose the vertices, which reproduce the leading order expressions of the four point blocks for $\langle VVVV \rangle$. However, for the next leading order in $1/c$, we also need the vertices, which reproduce the leading order expressions of the four point blocks for, say, $\langle WVWV \rangle$.
In this appendix, we compute the large $c$ limit of the four point blocks by following a technique explained for $\langle VVVV \rangle$ in section \ref{Basics}. See also \cite{Fitzpatrick:2014oza,Cornagliotto:2017dup} for an alternative approach with the Casimir equations.

We denote the operator product expansions schematically as
\begin{align}
W_{h_1} (z) V_{h_2} (0)= \sum_{p} z^{h_p - h_1 - h_2 - 1/2} \left( \tilde C_{12}^p [ \tilde V_{h_p} (0)]_\text{e} + C_{12}^p [\tilde V_{h_p} (0)]_\text{o} \right)
\label{ope2}
\end{align}
as in \eqref{ope}.
The superconformal families are denoted as $[\tilde V_h]_\text{e}$ and $[\tilde V_h]_\text{o}$, which include descendants with integer and half-integer levels, respectively.
We denote the superconformal families as
\begin{align}
[\tilde V_h (0)]_\text{e,o} = \sum_{N } z^N | \tilde N \rangle _{12}\, , \qquad |\tilde N \rangle _{12}  = \tilde C^{12}(N,h) V_h  \, , 
\end{align}
where the sums are over $N = 0,1,2 \ldots $ for $[\tilde V_h]_\text{e}$ and
$N = 1/2,3/2,5/2 \ldots $ for $[\tilde V_h]_\text{o}$.
The forms of the operators $\tilde C ^{12}(N , h)$  become simplified at the large $c$ limit as (see (4.24) of  \cite{Belavin:2007eq})
\begin{align}
\begin{aligned}
&\tilde 	C^{12}(n,h) = \frac{(h+ \tfrac12+h_1 - h_2 )_n}{n! (2 h)_{n}} G^{2n}_{-1/2} \, , \\
&\tilde 	C^{12}(n+ \tfrac12,h) = \frac{(h+h_1 - h_2 )_{n+1}}{n! (2 h)_{n+1}} G^{2n+1}_{-1/2} 
\end{aligned}
\label{CNh}
\end{align}
for non-negative integer $n$.

The operator product expansions \eqref{ope2} lead to the decomposition
\begin{align}
\begin{aligned}
&\langle W_{h_1} (z) V_{h_2} (0) W_{h_3} (\infty) V_{h_4} (1) \rangle  \\
& \qquad = \sum_{p} \tilde C^p_{12} \tilde C^{p}_{34} |\tilde F_\text{e}  (h_i ; h_p ;z)|^2 +  
\sum_{p} C^p_{12} C^{p}_{34} |\tilde F_\text{o} (h_i; h_p ;z)|^2 \, ,
\end{aligned}
\end{align}
where

\begin{align}
\begin{aligned}
&\tilde F_\text{e} (h_i;h_p;z) = z^{h_p - h_1 - h_2-1/2} \sum_{n=0}^\infty z^n  {}_{12}\langle \tilde{ n }  | \tilde{ n }  \rangle  _{43} \, , \\
&\tilde F_\text{o} (h_i;h_p;z) = z^{h - h_1 - h_2-1/2} \sum_{n=0}^\infty z^{n+1/2}  {}_{12} \langle \widetilde{ n + \tfrac12 } | \widetilde{ n + \tfrac12 }\rangle _{43} \, .
\end{aligned}
\end{align}
With \eqref{CNh} and  \eqref{GGV},
we can find the large $c$ limit of superconformal blocks as
\begin{align}
&\tilde F_\text{e} (h_i;h_p;z) = z^{h_p - h_1 - h_2- 1/2} {}_2 F_1 (h_p + h_1 - h_2 +\tfrac12 , h_p + h_4 - h_3 + \tfrac12 ; 2h_p ;z) + \mathcal{O}(c^{-1}) \nonumber \, , \\
&\tilde F_\text{o} (h_i;h_p;z)  = \frac{(h_p +h_1 -h_2)(h_p +h_4 -h_3) z^{h_p - h_1 - h_2}}{2h_p}\label{global}
\\
& \qquad \qquad \qquad \qquad \qquad  \times  {}_2 F_1 (h_p + h_1 - h_2 +1, h_p + h_4 - h_3 +1 ; 2h_p +1 ;z) + \mathcal{O}(c^{-1})\, . \nonumber
\end{align}

\subsection{Superconformal blocks with degenerate operators}
\label{CFT}

It is generically difficult to compute conformal blocks with primary operators.
However, if one of the operators is a degenerate one, then we may be able to obtain the expressions  by solving differential equations.
Here we consider a four point function with a simplest degenerate operator $V_{h_{3,1}}$, see \eqref{hmn}. 
We denote the four point function as
\begin{align}
g(z) = \langle V_{h_{3,1}} (z)  V_{h_{2}} (0) V_{h_{3}} (\infty) V_{h_{4}}  (1) \rangle \, .
\end{align}
Then, the differential equation for $g$ was obtained as \cite{Belavin:2006zr,Belavin:2007gz}
\begin{align}
\label{diff}
&\frac{1}{b^2} g''' + \frac{1 - 2 b^2}{b^2} \frac{1 - 2 z}{z(1 - z)} g '' + \left( \frac{b^2 + 2 h_2}{z^2} + \frac{b^2 + 2 h_4}{(1 - z)^2} + \frac{2 - 3 b^2 + 2 h_{24}}{z (1-z)} \right ) g'  \\
&+ \left( \frac{2 h_4 (1 + b^2)}{(1- z)^3} - \frac{2 h_2 (1 + b^2)}{z^3} + \frac{h_4 - h_2 + (1 - 2 z) (b^4 + b^2 (\frac12 - h_{24}) - h_2 - h_4 )}{z^2 (1-z^2)}\right) g = 0
\nonumber
\end{align}
with $h_{24} = h_2 + h_4 - h_3$.
The parameter $b$ is related to the central charge $c$ as%
\footnote{The parameter $Q$ can be identified as the background charge of the $\mathcal{N}=1$ super Liouville field theory.}
\begin{align}
c = \frac{3}{2}+ 3 Q ^2 \, , \quad Q = b + \frac{1}{b} \, ,
\end{align}
and it can be expanded in $1/c$ as
\begin{align}
b =\sqrt{3} \left (  \frac{1}{c} \right)^{1/2}  +\frac{15}{4} \sqrt{3}  \left(\frac{1}{c}\right)^{3/2}+ \mathcal{O}(c^{-5/2}) \, .
\end{align}
There are several choices of branch, but they are all equivalent.

We need the superconformal blocks for the four point function of the type in \eqref{4pt0}, thus we set $h_2 = h_{3,1}$ and $h_3 = h_4 = h_j \, (\equiv h_{4j+1,1}) $.
Since the differential equation is of the third order, there are three independent solutions, such as,
\begin{align}
\begin{aligned}
&z^{- 1- 2b^2}\mathcal{F}_\text{e}^{(+)} (z) =  1 + \sum_{n=1}^\infty a_{\text{e},n}^{(+)} z^n \, , \\
&z^{b^2} \mathcal{F}_\text{e}^{(-)} (z) = 1  + \sum_{n=1}^\infty a_{\text{e},n}^{(-)} z^n \, , \\
&z^{-1-b^2}\mathcal{F}_\text{o}^{(0)} (z) =  \frac{1}{2h_2} + \sum_{n=1}^\infty a_{\text{o},n}^{(0)} z^n \, .
\end{aligned}
\end{align}
The first one corresponds to the superconformal block with the exchange of identity operator.
Solving the differential equation \eqref{diff}, we obtain
\begin{align}
\label{iddeg}
& z^{- 1- 2b^2} \mathcal{F}_\text{e}^{(+)} (z) = 1 + \frac{1}{c} 
\biggl[ j z^2 + j z^3 + \frac{9j}{10} z^4 + \frac{4j}{5} z^5 + \frac{5j}{7} z^6 + \frac{9j}{14} z^7 + \frac{7j}{12} z^8 \biggr] \\
& \quad + \frac{1}{c^2} 
\biggl[ \left( 6 j^2+9 j \right) z^2 + \left( 6 j^2+9 j\right)  z^3 + \left( \frac{57 j^2}{10}+\frac{1611 j}{200}\right) z^4 + \left( \frac{27 j^2}{5}+\frac{711 j}{100}\right) z^5
\nonumber \\
&\quad  + \left(\frac{359 j^2}{70}+\frac{6177 j}{980} \right)  z^6 + \left(\frac{171 j^2}{35}+\frac{11043 j}{1960} \right)  z^7 + \left( \frac{1307 j^2}{280}+\frac{17069 j}{3360}\right)  z^8 \biggr] + 
\cdots  \nonumber 
\end{align}
up to the orders $1/c^2$ and $z^8$.
The second and third ones are the superconformal blocks with the descendant levels of exchanged operators are  non-negative integer and positive half-integer, respectively.
From \eqref{diff} we find 
\begin{align}
\begin{aligned}
& z ^{b^2}  \mathcal{F}_\text{e}^{(-)} (z) = 1 - \frac{1}{2}z + \frac{1}{c} \biggl[- \frac92 z + \left(3 j+\frac{9}{4} \right) z^2 + \left(\frac{3 j}{2}+\frac{3}{4} \right) z^3  + \left(j+\frac{3}{8}\right) z^4 \\ 
& \quad + \left(\frac{3 j}{4}+\frac{9}{40} \right) z^5   + \left( \frac{3 j}{5}+\frac{3}{20}\right) z^6   + \left( \frac{j}{2}+\frac{3}{28}\right) z^7   + \left( \frac{3 j}{7}+\frac{9}{112}\right) z^8  \biggr] + \cdots \, ,\\
& 2 h_2 z^{-1-b^2} \mathcal{F}_\text{o}^{(0)} (z) = 1  + \frac{1}{c} \biggl[- \frac32 z + \left( j-\frac{1}{2} \right) z^2 + \left(j-\frac{1}{4}\right) z^3  + \left( \frac{9 j}{10}-\frac{3}{20}\right) z^4 \\ 
& \quad + \left(\frac{4 j}{5}-\frac{1}{10} \right) z^5   + \left( \frac{5 j}{7}-\frac{1}{14} \right) z^6   + \left(\frac{9 j}{14}-\frac{3}{56}\right) z^7   + \left(\frac{7 j}{12}-\frac{1}{24}\right) z^8      \biggr] + \cdots
\end{aligned}
\label{gendeg}
\end{align}
up to the orders $1/c$ and $z^8$.

We also compute a heavy-heavy-light-light block in \eqref{gs}
with odd $s$ by solving the same differential equation \eqref{diff}. For the $z \to 0$ channel, there is only the identity four point block as
\begin{align}
z^{- 1 - 2 b^2} \mathcal{G}_s(z) = 1 + \sum_{n = 1}^\infty a_{s,n} z^n \, .
\end{align}
In the large $c$ expansion, it is given as
\begin{align}
\begin{split}
&z^{- 1 - 2 b^2} \mathcal{G}_s (z) = 1+\left(\frac{s^2}{24}-\frac{1}{24}\right) z^2+\left(\frac{s^2}{24}-\frac{1}{24}\right) z^3+\left(\frac{s^4}{1920}+\frac{7 s^2}{192}-\frac{71}{1920}\right) z^4 \\ & +\left(\frac{s^4}{960}+\frac{s^2}{32}-\frac{31}{960}\right) z^5+\left(\frac{s^6}{322560}+\frac{67 s^4}{46080}+\frac{1237 s^2}{46080}-\frac{3043}{107520}\right) z^6 \\ & +\left(\frac{s^6}{107520}+\frac{9 s^4}{5120}+\frac{119 s^2}{5120}-\frac{2689}{107520}\right) z^7  +  \frac{1}{c} u_{s} (z) + \mathcal{O}(z^8 , c^{-2}) \, ,
\end{split}
\label{gsleading}
\end{align}
where the $1/c$ correction is
\begin{align}
\label{gsnext}
&u_s (z)  =\left(-\frac{s^2}{16}+\frac{s}{4}-\frac{3}{16}\right) z^2+  \left(-\frac{s^2}{16}+\frac{s}{4}-\frac{3}{16}\right) z^3 \\ & + \left(-\frac{17 s^4}{12800}+\frac{s^3}{160}-\frac{79 s^2}{1280}+\frac{7 s}{32}-\frac{2073}{12800}\right) z^4+ \left(-\frac{17 s^4}{6400}+\frac{s^3}{80}-\frac{39 s^2}{640}+\frac{3 s}{16}-\frac{873}{6400}\right) z^5\nonumber \\ &  +\left(-\frac{467 s^6}{45158400}+\frac{s^5}{17920}-\frac{24449 s^4}{6451200}+\frac{67 s^3}{3840}-\frac{385919 s^2}{6451200}+\frac{1237 s}{7680}-\frac{576773}{5017600}\right) z^6\nonumber  \\ & +\left(-\frac{467 s^6}{15052800}+\frac{3 s^5}{17920}-\frac{10169 s^4}{2150400}+\frac{27 s^3}{1280}-\frac{125519 s^2}{2150400}+\frac{357 s}{2560}-\frac{489639}{5017600}\right) z^7 \nonumber 
\end{align}
up to the order of $z^7$.

It will be useful to recall a heavy-heavy-light-light block computed for the bosonic case as (see \cite{Hijano:2013fja,Hikida:2018dxe})
\begin{align}
\begin{aligned}
\mathcal{G}^B_s (z) &= \langle \mathcal{O}_{2,1} (z) \mathcal{O}_{2,1} (0) \mathcal{O}_{1,s} (\infty) \mathcal{O}_{1,s} (1) \rangle \\
& = z^{- \frac{2 ((k_B+4)^2 -1)}{4 (k_B + 2) (k_B+3)} } (1 -z)^{-\frac{s-1}{2}} 
{}_2 F_1 \left( - s+1 , - \frac{1}{k_B+2} \, - \frac{2}{k_B+2 } , z \right)  \, .
\end{aligned}
\label{Gs}
\end{align}
Here $\mathcal{O}_{m,n}$ is the degenerate operator with label $(m,n)$ in the Virasoro minimal model, see, e.g., \cite{DiFrancesco:1997nk}.
The label $k_B$ is related to the central charge $c$, and it can be expanded in $1/c$ as 
\begin{align}
k_B = - 3 + \frac{6}{c} + \mathcal{O} (c^{-2}) \, .
\end{align}
Using this, we can expand \eqref{Gs} in $1/c$ as
\begin{align}
\begin{aligned}
&z^{-1 - 9/c} \mathcal{G}^B_s (z) \\
&= \frac{\left(1 - (1-z)^s\right) (1-z)^{\frac{1-s}{2}}}{s z}- \frac{6}{c} (1-z)^{\frac{1-s}{2}} \sum _{l=1}^{s-1} \frac{(-1)^l \left(H_l-\frac{2 l}{l+1}\right) z^{l} \Gamma (s)}{\Gamma (l+2) \Gamma (s-l)} +\mathcal{O}(c^{-2})
\end{aligned}
\label{Gsexp}
\end{align}
with the harmonic number $H_l = \sum_{m=1}^l \frac{1}{m}$.
We can show that the leading order term of \eqref{gsleading} in $1/c$ coincides with that of \eqref{Gs} or \eqref{Gsexp}.


\providecommand{\href}[2]{#2}\begingroup\raggedright\endgroup

\end{document}